\documentclass[lettersize,journal]{IEEEtran}
\usepackage{stfloats}
\usepackage{url}
\usepackage{verbatim}
\usepackage{cite}
\usepackage{amsmath,amssymb,amsfonts}
\usepackage{graphicx}
\usepackage{textcomp}
\usepackage{xcolor}
\usepackage{amsthm}
\usepackage{bm}
\usepackage{graphicx}
\usepackage{float} 
\usepackage{subfigure}
\usepackage{setspace} 

\usepackage{caption}

\usepackage[linesnumbered, ruled]{algorithm2e}
\usepackage{algpseudocode}
\usepackage{ragged2e}

\newtheorem{myDef}{Definition}

\begin{document}

\title{
QoE-based Semantic-Aware Resource Allocation for Multi-Task Networks
}

\author{\IEEEauthorblockN{Lei Yan, Zhijin Qin,~\IEEEmembership{Senior Member,~IEEE}, Chunfeng Li, Rui Zhang*,~\IEEEmembership{Member,~IEEE}, Yongzhao Li*,~\IEEEmembership{Senior Member,~IEEE}, and Xiaoming Tao,~\IEEEmembership{Senior Member,~IEEE}
}
\vspace{-10pt}
\thanks{This work was supported in part by the Shaanxi Provincial Key Research and Development Program under Grant 2022ZDLGY05-03, Grant 2022ZDLGY05-04, and Grant 2023-ZDLGY-33, and in part by the National Natural Science Foundation of China under Grant 62293484 and Grant 61925105. The conference version of this work has been presented at the IEEE Global Communications Conference 2022 \cite{OurGlobecom}. \textit{(Corresponding author: Rui Zhang; Yongzhao Li.)} }

\thanks{Lei Yan, Rui Zhang, and Yongzhao Li are with the School of Telecommunications Engineering, Xidian University, Xi’an 710071, China. Lei Yan is also with the 54th Research Institute of CETC, Shijiazhuang, Hebei 050081, China (e-mail: lyan@stu.xidian.edu.cn; rz@xidian.edu.cn; yzhli@xidian.edu.cn)

Zhijin Qin and Xiaoming Tao are with the Department of Electronic Engineering, Tsinghua University, Beijing 100084, China. They are also with the State Key Laboratory of Space Network and Communications, Beijing 100084, China, and the Beijing National Research Center for Information Science and Technology, Beijing 100084, China. (e-mail: qinzhijin@tsinghua.edu.cn; taoxm@mail.tsinghua.edu.cn). 

Chunfeng Li is with the 54th Research Institute of CETC, Shijiazhuang, Hebei 050081, China (e-mail: Lee.chunfeng@foxmail.com).

}
}
\maketitle
\vspace{-10pt}
\begin{abstract}
By transmitting task-related information only, semantic communications yield significant performance gains over conventional communications. However, the lack of mature semantic theory about semantic information quantification and performance evaluation makes it challenging to perform resource allocation for semantic communications, especially when multiple tasks coexist in the network. To cope with this challenge, we propose a quality-of-experience (QoE) based semantic-aware resource allocation method for multi-task networks in this paper. 
First, semantic entropy is defined to quantify the semantic information for different tasks, and the relationship between semantic entropy and Shannon entropy is analyzed.
Then, we develop a novel QoE model to formulate the semantic-aware resource allocation in terms of semantic compression, channel assignment, and transmit power. The compatibility of the formulated problem with conventional communications is further demonstrated.
To solve this problem, we decouple it into two subproblems and solved them by a developed deep Q-network (DQN) based method and a proposed low-complexity matching algorithm, respectively.
Finally, simulation results validate the effectiveness and superiority of the proposed method, as well as its compatibility with conventional communications.
\end{abstract}


\vspace{-10pt}
\section{Introduction}
In the past decades, wireless communications focused on the engineering problem, i.e., accurate and efficient symbol transmission, and have made remarkable progress on approaching Shannon capacity limit. However, the explosive growth of data traffic motivates a revolutionary communication paradigm with higher communication efficiency\cite{ParadigmShift}. By focusing on accomplishing specific tasks, semantic communications have been identified as a promising technique to break through the bottleneck of conventional communications\cite{NineChalenge6G,zhijin2021challenges}. Recently, many different semantic communication systems have been designed for different types of tasks, yielding great performance gains over conventional communications in transmission reliability and efficiency \cite{gold2018,DeepSC,JSCC-image}. In this context, how to allocate resources at the semantic level with the aim of optimizing the semantic information efficiency is a significant challenge\cite{semantic-empowered}. The issue of resource allocation turns more complicated if the network is with multiple tasks and multiple users of semantic and conventional communications.

\vspace{-5pt}
\subsection{Related Works}
There have been a plenty of semantic communication systems designed for different tasks. On this basis, resource allocation for such a semantic-aware network has attracted extensive attention. However, the lack of mature semantic theory, especially about semantic information quantification, is a major obstacle for this issue. In the following, we will review the related works on semantic communication system design, semantic-aware resource allocation, and semantic information quantification and performance evaluation.

\subsubsection{Semantic communication system design} A lot of semantic communication systems have been proposed recently. They are specific to particular tasks with different sources and task requirements. For the text transmission task, knowledge-graph based \cite{shiQoE} and deep learning (DL) based\cite{DeepSC} semantic communication systems were investigated, in which the semantic features of text are extracted through knowledge graph and DL techniques, respectively, transmitted over the air, and received to recover the meaning of the source at the receiver. In a similar way, semantic communication systems for the transmission of other types of sources, like image\cite{JSCC-image, Image-semantic}, speech\cite{DeepSC-S}, and video\cite{2022vedio, vedio}, have shown better performance than conventional communications. In addition, intelligent tasks to be executed at the receiver have been paid attention as well, such as image retrieval \cite{SingleModalImage}, image classification\cite{ImageClassification}, and speech recognition\cite{SpeechRecognition}. The semantic communication systems for these tasks also achieve higher transmission efficiency than conventional ones since only task-related information are transmitted.

The aforementioned semantic communication systems are designed for single-modal tasks where a single user transmits single-modal data. For multi-modal tasks, multiple users transmit different modalities of data, and the received signals will be fused at the receiver to serve the task. In this case, multiple users jointly decide the task performance, which brings challenges to the system design as well as the resource allocation. Xie \textit{et. al.} took the visual question answering (VQA) task as an example, and then proposed a unified framework, named DeepSC-VQA, to support multi-modal tasks\cite{DeepSC-VQA}. In this system, the text and image from two users are fed into the Transformer based network, received at the receiver, and then fused though a layer-wise Transformer to predict the answer. 

Since semantic communication systems differ considerably for different tasks, the coexistence of multiple tasks should be considered for designing semantic-aware networks. Consequently, it should be taken into account for investigating the semantic-aware resource allocation strategy.

\subsubsection{Semantic-aware resource allocation}
There have been several preliminary studies in this realm. In \cite{RAPerformanceOptimization} and \cite{SemanticCompressionPC}, the importance weights of the extracted semantic features were considered. In addition to typical optimization variables, the partial semantic information to be transmitted was optimized in \cite{RAPerformanceOptimization} for the text transmission task with a knowledge graph based framework. The compression ratio of semantic features was investigated in \cite{SemanticCompressionPC} for the image classification task. However, both works aimed to maximize the task performance, without evaluating the semantic communication efficiency. Although Xia \textit{et al.} \cite{infocom} defined the system throughput for text transmission and formulated the user association and spectrum allocation problem to maximize the system throughput, they optimized the resource allocation at the message level rather than the semantic level.

To investigate the resource allocation at the semantic level, our prior work \cite{OurWork} defined the semantic transmission rate (S-R) and the semantic spectral efficiency (S-SE), and formulated a problem to maximize the overall S-SE. Similarly, the authors in \cite{DRL-driven} jointly optimized the compression ratio, transmit power, and bandwidth to maximize the long-term transmission efficiency. In addition, the works in \cite{Semantic-D2D} and \cite{Toward} studied the semantic-aware resource allocation in Device-to-Device vehicular networks and the network combining energy harvesting, cognitive radio, and non-orthogonal multiple access, respectively. However, none of these works considered the coexist of multiple multi-modal tasks or the compatibility of the method with conventional communications.

\subsubsection{Semantic information quantification and performance evaluation}
A communication system is usually evaluated from two aspects, accuracy and efficiency. The conventional communications are measured by bit-error rate and bit transmission rate. For semantic communications, the accuracy can be measured by the task performance and easily quantified by semantic similarity for text transmission, character-error-rate for speech recognition, answer accuracy for VQA task, and so forth. However, the efficiency of semantic communications is usually hard to measure and quantify. Although the S-SE has been defined in \cite{OurWork}, it can not be calculated due to the lack of semantic information quantification.

As early as 1952, the semantic information in a sentence was first measured based on the logical probability \cite{logic}. Similarly, focusing on the language system, a comprehension-centric semantic entropy was proposed and derived in terms of the structure of the world based on a comprehension model \cite{VenhuizenComprehension}. On the other hand, Melamed\cite{melamed1997measuring} measured the semantic entropy of the translation task through an information-theoretic method.  
Different from these definitions that are for a single task, Chattopadhyay \textit{et al.}\cite{chattopadhyay2020quantifying} proposed to quantify the semantic information for any task and type of sources with defining the semantic entropy as the minimum number of semantic queries about the data whose answers are sufficient to predict the answer. However, it is challenging to design the optimal semantic encoder to derive the semantic entropy. All aforementioned definitions indeed provide some insights on quantifying semantic information, but they failed to provide an operational method for practical implementation. Thus, a unified definition and a generic approach are still missing for deriving the semantic entropy.

Note that the semantic information quantification will contribute a lot to the performance evaluation, with which the S-R or S-SE for different tasks can be fairly measured. 
It may even provide a feasible way to evaluate the conventional communications at the semantic level, facilitating the semantic-aware resource allocation in a network where semantic and conventional systems coexist. In this context, more general performance metrics need to be developed to capture the characteristics of semantic communications as well as keep compatible with conventional communications. 

\vspace{-10pt}
\subsection{Motivations and Contributions}
The existing resource allocation schemes for semantic communications either did not consider the communication efficiency at the semantic level or focused on a single task. Moreover, the semantic information quantification, which plays a significant role on the performance evaluation of resource allocation, is still missing.  
On the other hand, the performance of semantic communications depends on whether the user can complete the task, focusing more on the subjective user experience. Therefore, semantic communications inherently have the advantage of enhancing quality-of-experience (QoE) \cite{shiQoE}, which considers both objective quality-of-service (QoS) and subjective user experience\cite{QoESemantic}. Compared with QoS-based resource allocation methods, QoE-based methods aim to improve the user satisfaction rather than the technical performance. Therefore, we study the semantic-aware resource allocation based on QoE to effectively and fairly allocate resources for users in the network where multiple modalities, multiple tasks, and multiple communication systems coexists.
The main contributions of this paper are as follows: 
\begin{itemize}
\item The semantic entropy is defined to quantify semantic information for the users with different tasks, and the relationship between semantic entropy and Shannon entropy is analyzed. Furthermore, a DL aided method is developed to obtain an approximate semantic entropy, which makes it possible to perform fair resource allocation in multi-task networks.
\item A semantic-aware QoE model is proposed in terms of semantic fidelity and semantic rate to better characterize user satisfaction, and then used to formulate the semantic-aware resource allocation with respect to semantic compression, channel assignment, and transmit power. Moreover, the compatibility of the formulated problem with conventional communications is demonstrated.
\item The formulated problem is decoupled into two subproblems, semantic compression subproblem and channel assignment and power allocation subproblem. With the semantic accuracy approximated by the deep neural network (DNN), a deep Q-learning (DQN) based method is developed to solve the first subproblem. Then the second one is modeled as a many-to-one
matching game and solved by a low-complexity matching algorithm.
\end{itemize}

Compared with the conference version \cite{OurGlobecom}, this journal version further clarifies the relationship between semantic entropy and Shannon entropy, analyzes the compatibility of the proposed method with conventional communications, develops an algorithm combining reinforcement learning and matching theory to provide a better solution, compares the proposed semantic-aware resource allocation model and the conventional one and validate the compatibility of the proposed method through simulations.

\vspace{-10pt}
\subsection{Organization and Notations}
The rest of this paper is organized as follows. Section II introduces the system model. Section III defines the semantic entropy and develops a novel QoE model to formulate the semantic-aware resource allocation problem. In Section IV, the optimization problem is solved by a solution with DQN and matching theory. Section V presents the simulation results, and Section VI concludes this paper.

\textit{Notation:} Bold-font variables represent matrices and vectors. Calligraphic-font variables represent sets. The superscript, $\mathbf{A}^{\rm{H}}$, represents the conjugate transpose of $\mathbf{A}$. $\mathbb{E}(x)$ is the expectation of $x$. ${\rm{dim}}({\rm{V}})$ represents the dimension of ${\rm{V}}$. In addition, $x\sim U(a,b)$ means following a uniform distribution over the interval $[a,b]$, and $x\sim N(\mu,\sigma^2)$ means following a normal distribution with mean $\mu$ and covariance $\sigma^2$.

\vspace{-5pt}
\section{System Model}

\begin{figure}
\vspace{-10pt}
  \centering
  \includegraphics[width=0.38\textwidth]{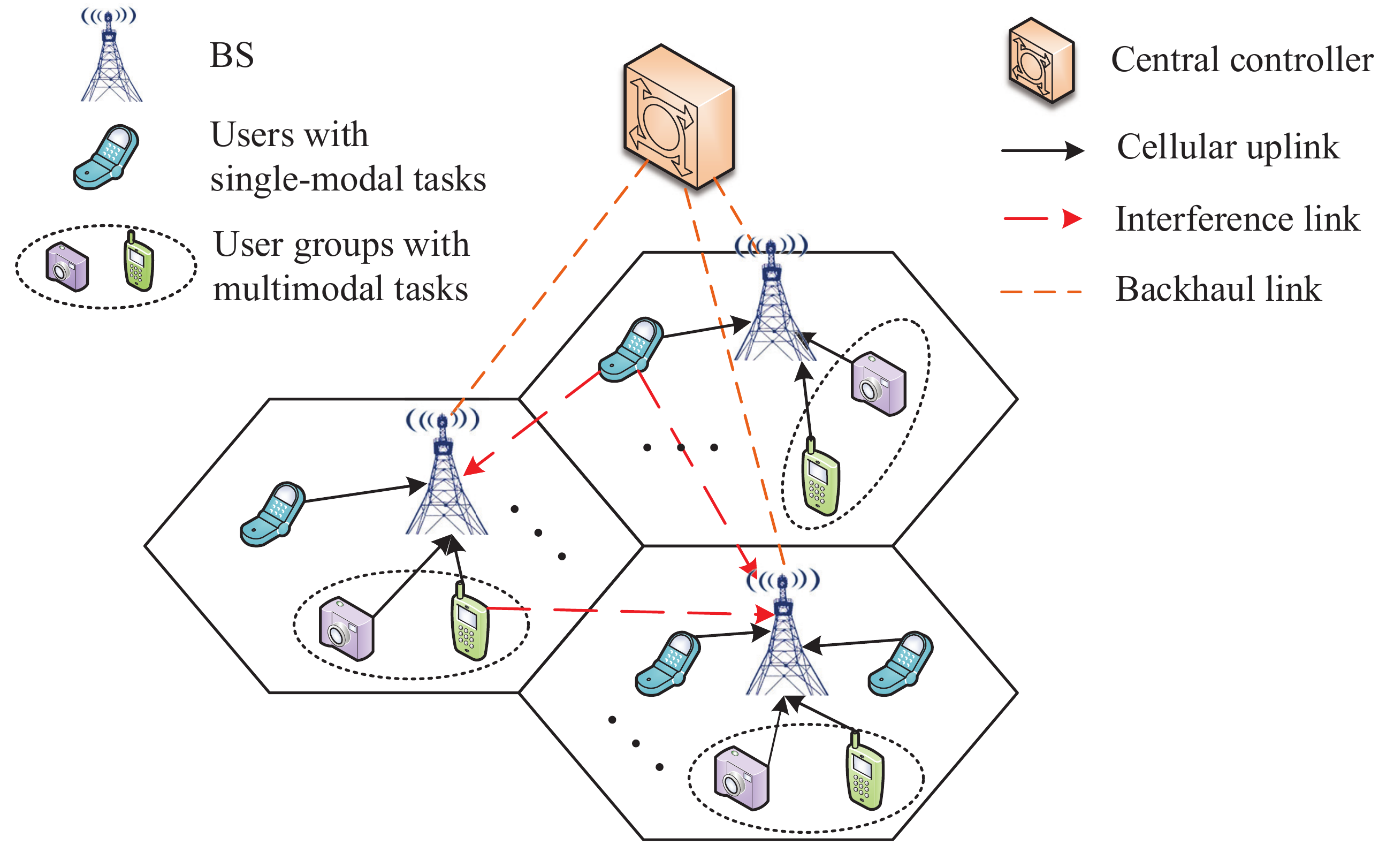}
  \caption{The multi-cell multi-task uplink cellular network.}
  \vspace{-15pt}
\end{figure}

As shown in Fig. 1, we consider an uplink cellular network with $B$ coordinated cells, where each cell has a base station (BS) with $N_{\rm{r}}$ receiving antennas and a set of single antenna users. Let $\mathcal{U}^b$ denote the user set in the $b$-th cell. We assume that $B$ BSs form a cooperation cluster, which can exchange channel state information (CSI) and scheduling information with a central controller through high-speed low-delay fiber backhaul links \cite{LuCooperation}. In addition, different intelligence tasks coexist in the network, including single-modal tasks and multi-modal tasks. Task-specific semantic communication transmitters and receivers are equipped at users and BSs, respectively. 
The transmission model and semantic communication models are introduced below.

\vspace{-10pt}
\subsection{Transmission Model}
In order to fully exploit spectrum resources, we assume that the $B$ coordinated cells share the same channel set denoted by  $\mathcal{M}$, and each user can occupy at most one channel with bandwidth $W$. In each cell, the channels are orthogonally allocated to users to eliminate the intra-cell interference. With the maximal ratio combining (MRC) detection at the BS, the receiving signal-to-interference-plus-noise ratio (SINR) of user $u, u\in \mathcal{U}^b$ at the $b$-th BS can be expressed as
\begin{equation}
    \gamma_u^b=\sum\limits_{m \in \mathcal{M}} \frac{\alpha_{u,m} p_{u} |{\bf{W}}_{u,m}^{b} {\bf{H}}_{u,m}^{b} |^2}{\|{\bf{W}}_{u,m}^{b} \|^2 \sigma^2+I_{u,m}^{b}},
\end{equation}
where $\alpha_{u,m}\in \{0,1\}$, $\alpha_{u,m}=1$ if channel $m \in \mathcal{M}$ is allocated to user $u$ and $\alpha_{u,m}=0$ otherwise, $p_{u}$ is the transmit power, and ${\bf{H}}_{u,m}^{b}$ is the $N_{\rm{r}} \times 1$ channel matrix from user $u$ to the $b$-th BS over channel $m$. Here, perfect CSI is assumed, which can be estimated through many channel estimation methods such as the linear least squares (LS) estimation.\footnote{Due to the page limitation, we do not simulate the impact of imperfect CSI in this paper. However, the authors in \cite{CSI-1,CSI-2} have explored the impact of imperfect CSI, i.e., as the CSI estimation error increases, the performance will decrease.} Accordingly, the MRC detection matrix can be obtained by ${\bf{W}}_{u,m}^{b}=({\bf{H}}_{u,m}^{b})^{\rm{H}}$. In Eq. (1), $\sigma^2$ is the noise power spectral density and $I_{u,m}^{b}$ is the interference experienced by user $u$ over channel $m$ from the adjacent cells, that is
\begin{equation}
    {I}_{u,m}^{b} = \sum\limits_{b'\in\mathcal{B},b'\ne b} \sum\limits_{u'\in \mathcal{U}^{b'}} \alpha_{u',m} p_{u'} |{\bf{W}}_{u,m}^{b} {\bf{H}}_{u',m}^{b} |^2.
\end{equation}

\vspace{-10pt}
\subsection{Semantic Communication Models}
We focus on two types of tasks in this paper, including a single-modal task and a bimodal task. However, the proposed algorithm can be extended to the case of multiple multi-modal tasks easily. 
Assume that $N_{\rm{Si}}$ single-modal users and $N_{\rm{Bi}}$ bimodal user pairs are randomly deployed in the considered network, and the $b$-th cell has $N_{\rm{Si}}^{b}$ single-modal users and $N_{\rm{Bi}}^b$ bimodal user pairs.
For ease of clarification, a single-modal user or a bimodal user pair is regarded as a user group. The index set of all user groups in the $b$-th cell is given by
\begin{equation}
\begin{split}
    {\mathcal{Q}}^b&=\{1,2,\dots,q,\dots,|{\mathcal{Q}}^b|\} \\
    &=\{1,2,\dots,N_{\rm{Bi}}^b,N_{\rm{Bi}}^b+1,\dots,N_{\rm{Bi}}^b+N_{\rm{Si}}^{b}\},
\end{split}
\end{equation}
where $|{\mathcal{Q}}^b|=N_{\rm{Bi}}^b+N_{\rm{Si}}^{b}$. Accordingly, the $q$-th user group contains a bimodal user pair when $q \le N_{\rm{Bi}}^b$, and it contains a single-modal user when $q > N_{\rm{Bi}}^b$. Besides, we denote the set of all user groups in the $b$-th cell by $\mathcal{G}^b=\{\mathcal{G}_q^b\}_{q\in \mathcal{Q}^b}$ where $\mathcal{G}_q^b \subseteq \mathcal{U}^b$ represents the user set of the $q$-th group.

For the single-modal task, we take text transmission task as an example. Specifically, DeepSC \cite{DeepSC} is adopted, where the sentence generated by the user is first compressed to semantic symbols that can be directly transmitted over the physical channel by the DeepSC transmitter, and then recovered by the DeepSC receiver at the BS. Moreover, semantic similarity is used to evaluate the task performance, which is a function of the average number of transmitted semantic symbols per word and SINR \cite{OurWork}. Thus, the semantic similarity of the user group $\mathcal{G}_q^b, q> N_{\rm{Bi}}^b$ can be expressed as 
\begin{equation}
\xi_{q}^b=f_{\rm{Si}}(k_{u_{\rm{s}}},\gamma_{u_{\rm{s}}}^b), {u_{\rm{s}}}\in \mathcal{G}_q^b,
\end{equation}
where $k_{u_{\rm{s}}}$ is the average number of transmitted semantic symbols of user with the single-modal task, ${u_{\rm{s}}}$.

For the bimodal task, we take VQA task as an example and adopt the DeepSC-VQA model \cite{DeepSC-VQA}. This task involves two users for text and image transmission, respectively. The two users first extract the semantic symbols from text and image through the DeepSC-VQA transmitter, respectively, and then send them to the BS. The received semantic symbols of text and image will be fused by the DeepSC-VQA receiver to predict the answer. As the two users jointly decide the task performance, the answer accuracy are modeled as a function with respect to the average number of transmitted semantic symbols per word for the text transmission user, the average number of transmitted semantic symbols per image for the image transmission user, and the SINR of the two users, i.e., 
\begin{equation}
\xi_{q}^b=f_{\rm{Bi}}(k_{u_{\rm{bt}}}, k_{u_{\rm{bi}}},\gamma_{u_{\rm{bt}}}^b,\gamma_{u_{\rm{bi}}}^b), u_{\rm{bt}},u_{\rm{bi}}\in \mathcal{G}_q^b,
\end{equation}
where $u_{\rm{bt}}$ and $u_{\rm{bi}}$ represent the user for text transmission and the one for image transmission, respectively.

\vspace{-5pt}
\section{QoE-based Semantic-Aware Resource Allocation}
In this section, we first define semantic entropy to quantify semantic information for different tasks, and then develop a semantic-aware QoE model to formulate the semantic-aware resource allocation. Finally, we demonstrate the compatibility of the proposed resource allocation model with conventional communications.

\vspace{-5pt}
\subsection{Semantic Entropy} 
A key challenge to evaluate the semantic communication efficiency is to quantify semantic information of the source data. In the following, we first elaborate semantic information and then provide a way to quantify it through semantic entropy.

Semantic information is regarded as the meaning of the source data underlying the concrete expression. More specifically, \textit{semantic information is the effective information contained in the source data for accomplishing a specific task}. This indicates that (i) semantic information relies not only on the source data but also on the specific task, which is significantly different from the information defined by Shannon, and (ii) semantic information is obtained by removing the redundant information irrelevant to the task from the source data. Consequently, the same data may contain different amounts of semantic information for different tasks. For example, an image contains much more semantic information for the image reconstruction task than the image classification task.

In this regard, semantic entropy is usually used to measure the semantic information, which should depend on the source data and the task. Inspired by Chattopadhyay \textit{et al.}~\cite{chattopadhyay2020quantifying}, we define the semantic entropy as following.
\begin{myDef}
Given semantic source $\mathcal{X}$, semantic entropy is defined as the minimum average number of semantic symbols about data $X \in \mathcal{X}$ that are sufficient to predict task $Y$, i.e.,
\begin{equation}
\begin{split}
    &H(X;Y)\buildrel \Delta \over =\mathop {\min }\limits_{E_{\rm{S}}} \mathbb{E}\left({\rm{dim}}(Code^{E_{\rm{S}}}(X))\right), E_{\rm{S}}\in \mathcal{E}_{\rm{S}} \\
    &{\rm{s.t.}}\quad P(Y|Code^{E_{\rm{S}}}(X))=P(Y|X),
\end{split}
\end{equation}
where $Code^{E_{\rm{S}}}(X)$ denotes the semantic symbol vector extracted from $X$ with the semantic encoder ${E_{\rm{S}}}$, $\mathcal{E}_{\rm{S}}$ is the set of semantic encoders, and $P(Y|X)$ is the conditional probability of achieving the goal of~ $Y$ given $X$. 
\end{myDef}

The semantic symbol in Definition 1 is an element of the output of the semantic encoder, and in our work, we regard a channel symbol as a semantic symbol. The constraint in Definition 1 implies that the defined semantic entropy is lossless, and the semantic entropy is actually defined as an expected value over the whole data set $\mathcal{X}$, i.e., the semantic entropy is a constant for the same task and dataset, which shares a similar philosophy as \cite{OurWork}. 
However, it is intractable to find an optimal semantic encoder, ${E_{\rm{S}}^*}$, to derive the semantic entropy \cite{ParadigmShift}. To obtain a measure that is both meaningful and manipulable for semantic communication systems, we utilize a well-designed DL model as the encoder to obtain an approximate semantic entropy for a task, which is 
\begin{equation}
\begin{split}
    &{\tilde H}(X;Y)\buildrel \Delta \over =\mathop {\min } \mathbb{E}\left({\rm{dim}}( Code^{E_{\rm{DL}}}(X) )\right) \\
    {\rm{s.t.}}&\quad P(Y|X)-P(Y|Code^{E_{\rm{DL}}}(X))<\varepsilon,
\end{split}
\end{equation}
where the constraint indicates that the task performance degradation can not exceed $\varepsilon$. From Eq. (7), the defined approximate semantic entropy is lossy. Note that the data granularity of the approximate semantic entropy depends on the granularity of the data that the DL model can deal with. For example, the DeepSC transmitter maps a sentence into a semantic symbol vector with $k$ semantic symbols representing a word, and thus the obtained approximate semantic entropy through DeepSC corresponds to a word as well.

According to the aforementioned method, the approximate semantic entropy of the considered tasks can be derived based the corresponding DL models. In particular, we first remove the channel models from DeepSC and DeepSC-VQA, then train them under different settings of the number of semantic symbols, and finally find the minimum number of semantic symbols that can guarantee a performance very close to the upper bound. Furthermore, we define the unit of the approximate semantic entropy as \textit{sut} as in \cite{OurWork}.

\textit{Remark:} Shannon entropy \cite{Shannon} and semantic entropy are the bounds of lossless data compression in different compression ways, respectively, one is based on the statistical knowledge of the source and measured in \emph{bits}, and the other is based on the semantic correlations and measured in \emph{suts}. Note that there still lacks a unified definition of \emph{suts}, and defining the number of \emph{suts} with the number of channel symbols in our work is just one possible way to quantify semantic information. Hence, if we assume the semantic entropy is derived by compressing data into bits, i.e., 1 sut = 1 bit, Shannon entropy and semantic entropy are in the same units but with different compression ways. In this context, when Shannon entropy, $H(X)$, derived through the optimal encoder that is based on the statistical knowledge, $E_{\rm{C}}^*$, is smaller than the value derived through all encoders that are based on the semantic correlations, we can roughly regard $E_{\rm{C}}^*$ as ``\emph{the optimal semantic encoder that is based on the statistical knowledge}'', i.e., $E_{\rm{C}}^*=E_{\rm{S}}^*$, and thus
\begin{equation}
H(X;Y)(\emph{bits})=H(X)(\emph{bits}).
\end{equation}
This may happen when there exists very little redundant information in the source for the task, such as digital data transmission with high reliability. Nevertheless, \textit{whatever for semantic or conventional communication systems, the semantic entropy exists and is a constant with the same measure for the same task and data}.

\vspace{-5pt}
\subsection{Semantic-Aware QoE}
We formulate the semantic-oriented QoE model by considering the accuracy and efficiency of semantic communications. Specifically, two objective metrics are focused on, \textit{semantic fidelity} and \textit{semantic rate}. Here, semantic fidelity measures the difference of the semantic information between the source data and the recovered data, which corresponds to the semantic similarity of the considered single-modal task and the answer accuracy of the considered bimodal task. The latter is defined as the amount of semantic information emitted to the transmission medium per second, measured in \textit{suts/s}. The semantic rate of the single-modal user ${u_{\rm{s}}}\in \mathcal{G}_q^b, q> N_{\rm{Bi}}^b$ is given as
\begin{equation}
    \varphi_{u_{\rm{s}}}=\frac{\tilde H_{\rm{Si}}}{k_{u_{\rm{s}}}/W},
\end{equation}
where $\tilde H_{\rm{Si}}$ represents the DeepSC\cite{DeepSC} based approximate semantic entropy. 
The semantic rate of bimodal users $u_{\rm{bt}}$ and $u_{\rm{bi}}$, $u_{\rm{bt}},u_{\rm{bi}}\in \mathcal{G}_q^b,q \le N_{\rm{Bi}}^b$ are expressed as
\begin{equation}
    \varphi_{u_{\rm{bt}}}=\frac{\tilde H_{\rm{Bi,t}}}{k_{u_{\rm{bt}}}/W}, 
\end{equation}
and
\begin{equation}
    \varphi_{u_{\rm{bi}}}=\frac{\tilde H_{\rm{Bi,i}}}{k_{u_{\rm{bi}}}/W},
\end{equation}
respectively, where $\tilde H_{\rm{Bi,t}}$ and $\tilde H_{\rm{Bi,i}}$ represent the DeepSC-VQA\cite{DeepSC-VQA} based approximate semantic entropy for the text and the image transmission users, respectively.

Note that the semantic rate is significantly different from the S-R that is defined as the amount of \textit{successfully} delivered semantic information per second in \cite{OurWork}. With $\Gamma_u$ denoting the S-R, we have 
\begin{equation}
\Gamma_u=\varphi_u\xi_q,
\end{equation}
i.e., the difference between them lies in that whether the semantic fidelity is taken into account. The S-R couples semantic fidelity and semantic rate. However, from the perspective of users, semantic fidelity and communication efficiency are different, and users may have their own preferences on them depending on the applications. For example, some users prefer high fidelity but are delay-tolerated while others may desire higher semantic rate but do not need a very high fidelity. In order to reflect QoE requirements of users more properly, we consider the two QoS metrics to formulate the QoE model as
\begin{equation}
\begin{split}
    QoE_q^b&=\sum\limits_{u\in{\mathcal{G}_q^b}}({{w_u}G_u^{\rm{R}}+(1-w_u)G_u^{\rm{A}}})\\
    &=\sum\limits_{u\in{\mathcal{G}_q^b}}({\frac{{w_u}}{1+e^{\beta_u(\varphi_u^{\rm{req}}-\varphi_u)}}+\frac{(1-w_u)}{1+e^{\lambda_u(\xi_u^{\rm{req}}-\xi_q^b)}}}),
\end{split}
\end{equation}
where $QoE_q^b$ is the QoE of the $q$-th user group in the $b$-th cell, $w_u$ and $(1-w_u)$ are the weights of semantic rate and semantic fidelity at user $u$, respectively, $G_u^{\rm{R}}$ and $G_u^{\rm{A}}$ are the scores of semantic rate and semantic fidelity at user $u$, respectively, and $\beta_u$ and $\lambda_u$ represent the growth rates of $G_u^{\rm{R}}$ and $G_u^{\rm{A}}$, respectively. Additionally, $\varphi_u^{\rm{req}}$ and $\xi_u^{\rm{req}}$ represent the minimum semantic rate and semantic fidelity to acquire 50\% of the scores, respectively. Due to the simplicity and generality, we choose the logistic function\cite{LogisticQoE} to model the correlation between QoE and QoS metrics while other QoE function types are applicable as well for the proposed method, such as the mean opinion score (MOS) based function\cite{MosQoE} or the exponential function\cite{ExponentialQoE}. Here, $QoE_q^b$, $G_u^{\rm{R}}$, and $G_u^{\rm{A}}$ are between 0 and 1.

\vspace{-5pt}
\subsection{Problem Formulation}
In this part, we formulate the semantic-aware resource allocation with the goal of maximizing the overall QoE of all users in terms of channel assignment, transmit power, and semantic compression. The semantic compression here represents the average number of transmitted semantic symbols. The optimization problem can be expressed as follows:
\begin{align}
    {\rm{\mathbf{(P0)}}}&\quad  \mathop{\max}\limits_{\{\{\alpha_{u,m}\},\{p_{u}\},\{k_{u}\}\}}\  {\sum\limits_{b \in \mathcal{B}}{\sum\limits_{q\in{\mathcal{Q}^b}}{QoE_q^b}}} \label{AA}\\
    {\rm{ s.t.}}\ &{\rm{ C_1 }}:\alpha _{u,m} \in \{0,1\},\  \forall u \in  {\mathcal{U}}^b,\ \!\forall m \!\in\! \mathcal{M},\ \forall b \!\in\! \mathcal{B} \tag{\ref{AA}{a}}, \label{AAa}\\
    &{\rm{ C_2 }}:\sum\limits_{u \in  {\mathcal{U}}^b} {{\alpha _{u,m}} \le 1,\ \forall m \in \mathcal{M}},\ \forall b \!\in\! \mathcal{B}\tag{\ref{AA}{b}}, \label{AAb}\\
    &{\rm{    C_3   }}:\sum\limits_{m\in \mathcal{M}}{{\alpha _{u,m}} \le 1,\ \forall u \in {\mathcal{U}}^b},\ \forall b \!\in\! \mathcal{B} \tag{\ref{AA}{c}}, \label{AAc}\\
    &{\rm{     C_4  }}:\sum\limits_{u\in \mathcal{G}_q^b,q \le N_{\rm{Bi}}^b}\!\!\!\!\!\!{{\alpha _{u,m}} \in \{0,2\},\ \forall m \in {\mathcal{M}}},\ \forall b \!\in\! \mathcal{B} \tag{\ref{AA}{d}}, \label{AAd}\\
    &{\rm{     C_5  }}:{k_u} \in {\mathcal{K}}_u,\ \forall u \in {\mathcal{U}}^b,\ \forall b \!\in\! \mathcal{B} \tag{\ref{AA}{e}}, \label{AAe}\\
    &{\rm{     C_6  }}: 0 \le p_{u} \le P_{\rm{max}},\ \forall u \in {\mathcal{U}}^b,\ \forall b \!\in\! \mathcal{B} \tag{\ref{AA}{f}}, \label{AAf}\\
    &{\rm{C_7}}:G_{u}^{\rm{R}}, G_{u}^{\rm{A}} \ge G_{\rm{th}},\ \forall u \in {\mathcal{U}}^b,\ \forall b \!\in\! \mathcal{B} \tag{\ref{AA}{g}} \label{AAg},
\end{align}
where $\rm{C_1}$ constrains the range of $\alpha_{u,m}$, $\rm{C_2}$ ensures the orthogonal channels for each user in a cell, $\rm{C_3}$ restricts each user to occupy at most one channel, $\rm{C_4}$ ensures that a bimodal user pair will be allocated no channel or two channels as only one channel assigned to one of them will lead to a failure, $\rm{C_5}$ specifies the range of the number of transmitted semantic symbols for each user, and  $\mathcal{K}_u=\mathcal{K}_{\rm{Si}}, \mathcal{K}_{\rm{Bi,t}}, {\rm{and}} \  \mathcal{K}_{\rm{Bi,i}}$ when $u$ is the user with the single-modal task, the user with the bimodal task for text transmission, and the user with the bimodal task for image transmission, respectively, $\rm{C_6}$ constrains the range of transmit power, and $\rm{C_7}$ limits the minimum required scores of semantic rate and semantic fidelity. Among these constraints, $\rm{C_1}$, $\rm{C_2}$, $\rm{C_3}$, and $\rm{C_6}$ are similar to the constraints in the conventional resource allocation model while $\rm{C_4}$, $\rm{C_5}$, and $\rm{C_7}$ are specific to semantic communications. Particularly, $\rm{C_4}$ is the most different one, which is for the bimodal task and makes it more difficult to solve ${\rm{\mathbf{(P0)}}}$.

\vspace{-5pt}
\subsection{Compatibility with Conventional Communication Systems}
The users in a network, whatever adopting the semantic communication technique or the conventional communication technique, may share the same resource pool for data transmission. Therefore, a resource allocation model that is compatible to both systems and can optimally allocate resources to all users is desirable. However, the requirements of the users with the two communication techniques are at distinct levels for the resource allocation, and thus the first thing to achieve the ``compatibility'' of the semantic-aware resource allocation is to evaluate the two communication techniques at the same level. In the following, we will elaborate how to carry out the fair performance evaluation for semantic and conventional communication systems from the QoS and QoE perspectives, based on which the compatibility of the proposed resource allocation model is clarified.

\subsubsection{QoS}
For QoS, we focus on the information transmission rate, i.e., the S-R \cite{OurWork} for semantic communications and the bit transmission rate for conventional communications. The first thing to fairly evaluate the performance of the two kinds of systems is to establish a transformation relation between S-R and bit transmission rate. An initial exploration has been made in \cite{OurWork} for text transmission task, that is 
\begin{equation}
{\Gamma_u}'=\frac{C_u}{\mu _u}\frac{I}{L},
\end{equation}
where ${\Gamma_u}'$ is the equivalent S-R of conventional communication systems, $C_u=W{\log _2}(1 + {\gamma _u})$ is the bit transmission rate, $\mu_u$ is the transforming factor revealing the compression ability of the source coding, and $I/L$ is the expected amount of semantic information per word. However, $I/L$ is not quantified in \cite{OurWork}. Although this will not affect the solution for a single task, the work in \cite{OurWork} cannot cope with the coexistence of multiple tasks. Nevertheless, with the derived semantic entropy, Eq. (15) can be rewritten as 
\begin{equation}
{\Gamma_u}'=\frac{C_u}{\mu _u}{ H_u}.
\end{equation}
Accordingly, the transformation relation between S-R and bit transmission rate as shown in Eq. (16) is generic for all tasks.

Thus, by changing the objective function to the sum S-R of all semantic and conventional users, along with the constraint $\rm{C}_7$ to the S-R constraint, ${\rm{\mathbf{(P0)}}}$ can be used to formulate the resource allocation problem for the network where semantic and conventional communications coexist.

\subsubsection{QoE} QoE measures how satisfied the users are with the current service (i.e., the QoS). The QoE model of conventional communications is usually based on the bit transmission rate\cite{QoESwap}, which can be expressed as 
\begin{equation}
    {QoE_q^b}'=\sum\nolimits_{u\in{\mathcal{G}_q^b}}{G_u^{\rm{C}}}    =\sum\nolimits_{u\in{\mathcal{G}_q^b}}{\frac{1}{1+e^{{\beta_u^{\rm{C}}}({C_u^{\rm{req}}}-{C_u})}}},
\end{equation}
where $G_u^{\rm{C}}$ is the score of the bit transmission rate at user $u$, $\beta_u^{\rm{C}}$ represents the growth rate of $G_u^{\rm{C}}$, and $C_u^{\rm{req}}$ is the minimum bit transmission rate to acquire 50\% of the score. Similar to the developed QoE model for semantic communications, the logistic function is adopted here.

Although semantic and conventional communications focus on different QoS metrics, their QoE is between 0 and 1. Thus, to some extend, QoE model provides a more intuitive way to deal with the coexistence of semantic and conventional communications, with no need to find a transformation relation between them. Therefore, ${\rm{\mathbf{(P0)}}}$ can achieve fair and optimal resource allocation for all users in the network where semantic and conventional communications coexist.

\vspace{-5pt}
\section{Proposed Semantic-Aware Resource Allocation Algorithms}
In this section, we first decouple ${\rm{\mathbf{(P0)}}}$ into two subproblems, including a semantic compression subproblem and a channel assignment and power allocation subproblem. Then, a DQN based method is developed to solve the first subproblem and a low-complexity matching algorithm is proposed to solve the second one.

\vspace{-5pt}
\subsection{Problem Decoupling}
The problem ${\rm{\mathbf{(P0)}}}$ is a combinatorial optimization problem, which generally yields extremely high computational complexity if using an exhaustive search method. Hence, an effective method with low complexity is desired. However, we face the following challenges:
\begin{itemize}
\item [1)]There exists no close-form expression on $\xi_q^b$ with respect to $\gamma_u^b$ and $k_u$, which means that convex optimization methods are not applicable. Moreover, since the discrete mapping tables are obtained by training the DL models of different tasks, the computing cost will be extremely high if we want more results with different $\gamma_u^b$ and $k_u$.
\item [2)]Co-channel interference caused by the users from other cells brings difficulties to the optimization of $\{\alpha_{u,m}\}$ and $\{p_u\}$.
\item [3)]The performance optimization of the users in a bimodal user pair are tightly coupled ($\rm{C}_4$), making it more intractable to solve ${\rm{\mathbf{(P0)}}}$.
\end{itemize}

To cope with these challenges, we first analyze the coupling relation among optimization variables and objective function of ${\rm{\mathbf{(P0)}}}$. From Eq. (1), $\{\alpha_{u,m}\}$ and $\{p_u\}$ jointly decide SINR $\gamma_u^b$. Then, $\gamma_u^b$ and $k_u$ jointly decide $\xi_q^b$ according to Eq. (4) and (5), and further determine $QoE_q^b$ according to Eq. (13). 
Therefore, ${\rm{\mathbf{(P0)}}}$ can be decomposed into 
two subproblems, a semantic compression subproblem and a channel assignment and power allocation subproblem, as 
\begin{align}
    {\rm{\mathbf{(P1)}}}\quad  &\mathop{\max}\limits_{\{k_{u}\},u\in \mathcal{G}_q^b}\  {{{QoE_q^b}}}\label{YY}\\
    &{\rm{ s.t.}}\ \  {\rm{C_5}}\ {\rm{and}}\ {\rm{C_7}} \nonumber
    \text{,}
\end{align}
and 
\begin{align}
    {\rm{\mathbf{(P2)}}}\quad & \mathop{\max}\limits_{\{\{\alpha_{u,m}\},\{p_{u}\}\}}\  {\sum\limits_{b \in \mathcal{B}}{\sum\limits_{q\in{\mathcal{Q}^b}}{QoE_q^{b}}}}\label{YY}\\
    &{\rm{ s.t.}}\ \  {\rm{C_1}},{\rm{C_2}},{\rm{C_3}},{\rm{C_4}},\ {\rm{and}}\ {\rm{C_6}} \nonumber
    \text{,}
\end{align}
respectively. 

\begin{figure}
  \centering
  \includegraphics[width=0.4\textwidth]{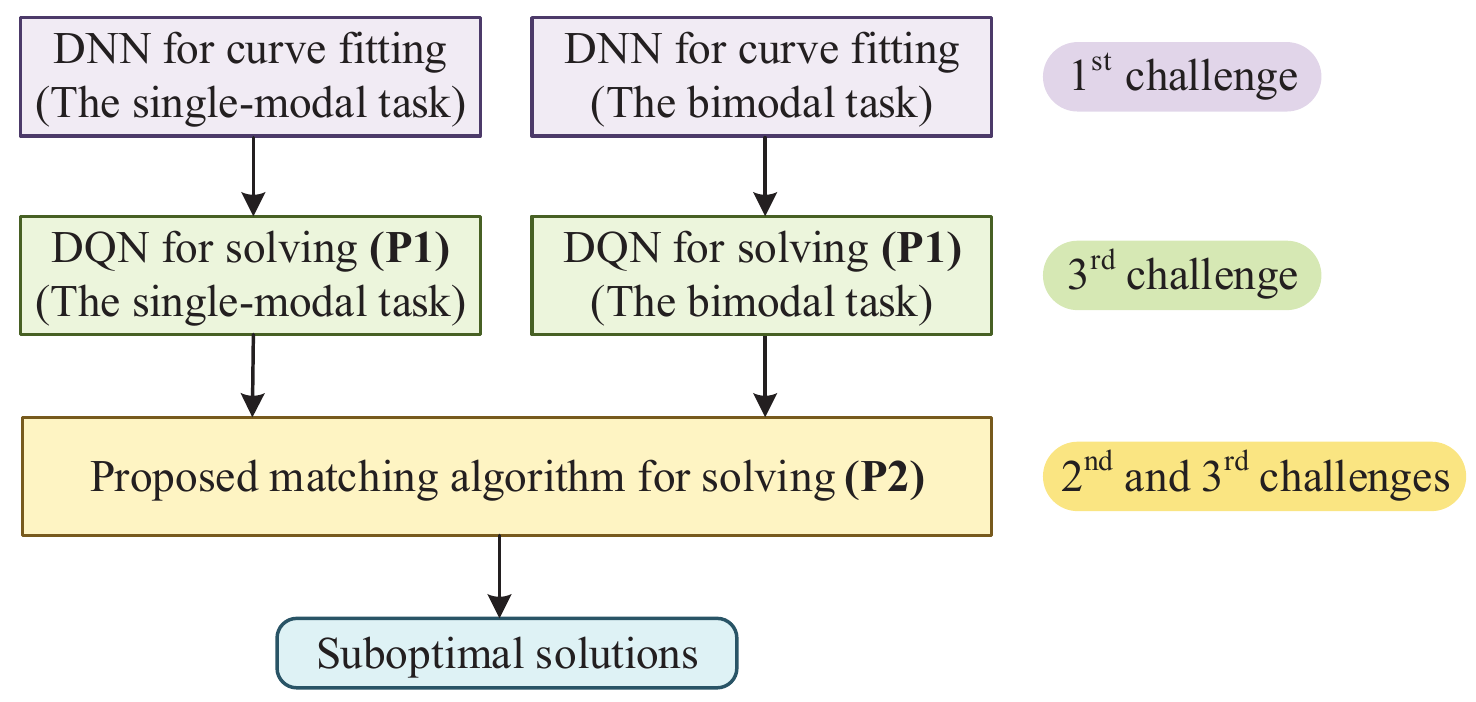}
  \caption{The framework of the proposed solution.}
  \label{solution}
  \vspace{-15pt}
\end{figure}

Fig.~\ref{solution} presents the framework of the proposed solution to address the above three challenges. We first perform curve fitting to predict the semantic fidelity of different tasks. Though some researchers have attempted to fit the semantic fidelity with a logistic function \cite{SemiNoma} or an exponential function \cite{SemanticCompressionPC}, they are for the single-modal tasks that are with fewer parameters as in Eq. (4). For multimodal tasks, it is hard to use these functions to fit the performance due to many more parameters of different users and the complicated relations among them. Considering that neural networks have good fitting ability, we adopt DNN to fit the semantic fidelity for each task. The fitting performance will be shown in Section V-A. With these trained DNNs, the semantic fidelity can be obtained at any $\gamma_u^b$ and $k_u$. Then, ${\rm{\mathbf{(P1)}}}$ is solved by the reinforcement learning (RL) method. Specifically, two DQNs are designed to find the suboptimal semantic compression, $k_{u_{\rm{s}}}^*$ or $(k_{u_{\rm{bt}}}^*,k_{u_{\rm{bi}}}^*)$, and the corresponding $QoE_q^{b*}$ for the two tasks, respectively. Finally, a low-complexity matching algorithm is developed for ${\rm{\mathbf{(P2)}}}$.

\vspace{-5pt}
\subsection{Semantic compression subproblem}
Since we have to solve ${\rm{\mathbf{(P1)}}}$ for every user group in every possible matching during the process for solving ${\rm{\mathbf{(P2)}}}$, using the exhaustive search to solve ${\rm{\mathbf{(P1)}}}$ as in \cite{OurGlobecom} will lead to high computation complexity. To cope with this problem, we model ${\rm{\mathbf{(P1)}}}$ as a RL problem, and develop a DQN based method to learn an optimal approximated policy to maximize the QoE for each user group. The components of the DQN for the single-modal task are as follows:
\begin{itemize}
\item \textit{Agent}: The agent is the central controller that performs the centralized resource allocation for all users in the cooperative cell cluster. 
\item \textit{States}: We take SINR $\gamma_{u_{\rm{s}}}^b$ and other QoE-related parameters as the states to enable the DQN adapt to different QoE requirements and SINR. Thus, the state space is defined as ${\mathcal{S}}_{\rm{Si}}=\{s_t,s_t=\{\gamma_{u_{\rm{s}}}^b, w_{u_{\rm{s}}},\beta_{u_{\rm{s}}},\varphi_{u_{\rm{s}}}^{\rm{req}}, \lambda_{u_{\rm{s}}}, \xi_{u_{\rm{s}}}^{\rm{req}}, G_{\rm{th}}\}\}$.
\item \textit{Actions}: The action space consists of all possible values of $k_{u_{\rm{s}}}, {u_{\rm{s}}}\in \mathcal{G}_q^b, q> N_{\rm{Bi}}^b$, i.e., ${\mathcal{A}}_{\rm{Si}}={\mathcal{K}}_{u_{\rm{s}}}$, which satisfies constraint $\rm{C_5}$.
\item \textit{Reward}: At the $t$-th training step, the reward of choosing action $a_t \in {\mathcal{A}}_{\rm{Si}}$ at state $s_t \in{\mathcal{S}}_{\rm{Si}}$ is 
\begin{equation}
r_t=\left\{
    \begin{split}
        &QoE_q^b,q> N_{\rm{Bi}}^b,\quad {\rm{if}} \ {\rm{C_7}}\  {\rm{is}}\ {\rm{satisfied}};\\
        &0, \quad \quad\quad \quad\quad\quad\quad {\rm{otherwise}}.
    \end{split}
    \right.
\end{equation}
Then the expected discount reward can be expressed as
\begin{equation}
{R_t} = \sum\nolimits_{\tau = 0}^\infty {\alpha^\tau {r_{t+\tau}}},
\end{equation}
where $\alpha \in [0,1]$ is the discount rate to determine the weight of the future reward.

\item \textit{Policy}: The $\epsilon$-greedy policy is applied to make a trade-off between exploitation and exploration for choosing actions. The agent will select an action randomly from ${\mathcal{A}}_{\rm{Si}}$ with the probability of $1-\epsilon$, while choosing the action with the maximum Q-value with the probability $\epsilon$.
\end{itemize}

Similarly, the components of the DQN for the bimodal task are as follows:
\begin{itemize}
\item \textit{Agent}: The agent is the central controller as well.
\item \textit{States}: The state space of the bimodal task includes the states of two users, which is defined as ${\mathcal{S}}_{\rm{Bi}}=\{s_t, s_t=\{\gamma_{u_{\rm{bt}}}^b,w_{u_{\rm{bt}}}, \beta_{u_{\rm{bt}}}, \varphi_{u_{\rm{bt}}}^{\rm{req}}, \lambda_{u_{\rm{bt}}}, \xi_{u_{\rm{bt}}}^{\rm{req}}, \gamma_{u_{\rm{bi}}}^b, w_{u_{\rm{bi}}},\beta_{u_{\rm{bi}}},\varphi_{u_{\rm{bi}}}^{\rm{req}},$ $ \lambda_{u_{\rm{bi}}}, \xi_{u_{\rm{bi}}}^{\rm{req}}, G_{\rm{th}}\}\}$.

\item \textit{Actions}: The actions include all possible combinations of the number of semantic symbols of two users, i.e., ${\mathcal{A}}_{\rm{Bi}}=\{(k_{u_{\rm{bt}}},k_{u_{\rm{bi}}}),\forall k_{u_{\rm{bt}}} \in {\mathcal{K}}_{u_{\rm{bt}}}, \forall k_{u_{\rm{bi}}} \in {\mathcal{K}}_{u_{\rm{bi}}}\}$.

\item \textit{Reward}: The reward function is similar to that for the single-modal task but $q \le N_{\rm{Bi}}^b$.

\item \textit{Policy}: The $\epsilon$-greedy policy is adopted as well.
\end{itemize}

After illustrating the components of the DQNs for the two considered tasks, the detailed training process will be introduced in the following. In the DQN based RL approach, there are two fully connected DNNs for approximating the state-action value function $Q(s_t, a_t)$, the evaluation network and the target network. The inputs of both networks are $s_t$, and the outputs of the evaluation network and the target network are the estimated Q-values and the target Q-values of all actions, respectively. The weights of the target network, $\bm{\bar \theta}$, will be updated with the evaluation network weights, ${\bm{\theta}}$, after every $I$ iterations. To train the evaluation network, the transition $e_t=(s_t, a_t,r_t,s_{t+1})$ at every training step will be collected in a memory buffer $\mathcal{D}$, where a minibatch of this set is randomly sampled and acts as the training data. The loss function of the evaluation network is 
\begin{equation}
Loss({\bm{\theta}})=\mathbb{E}[(r_t+\alpha \mathop{\max}\limits_{a_{t+1}}Q(s_{t+1}, a_{t+1}|\bm{\bar \theta})-Q(s_{t}, a_{t}|\bm{\theta}))^2 ],
\end{equation}
based on which ${\bm{\theta}}$ is optimized by the gradient descent method.

\subsection{Channel Assignment and Power Allocation Subproblem}
To cope with the tight coupling among users in multiple cells and those in a bimodal user pair, we construct a matching game to model ${\rm{\mathbf{(P2)}}}$ and propose a low-complexity matching algorithm to obtain the stable matching in this part.

Problem ${\rm{\mathbf{(P2)}}}$ is a three-sided many-to-one matching game among users, channels, and power levels, where each user can select at most one channel and one  power level while each channel or power level can serve multiple users. For the convenience of algorithm implementation, we put all combinations of channels ($\mathcal{M}$) and power levels ($\mathcal{P}$) together to form a resource set $\mathcal{T}\!\!=\!\!\{(m,p),\forall m \!\in\! \mathcal{M}, \forall p \!\in\! \mathcal{P}\}$. Then the three-sided matching can be converted to a two-sided matching. In addition, the QoE of each user depends not only on the opposite partners to be matched but also on the users of other cells sharing the same channel. More specifically, this is a matching problem with externalities. The idea of swap matching \cite{SwapMatching} can be used for reference to obtain the solution.

Since two cases may exist in each cell, i.e., $|\mathcal{U}^b|\le|\mathcal{M}|$ and $|\mathcal{U}^b|>|\mathcal{M}|$, we focus on the markets of different sides to maximize the overall QoE of the two cases, respectively.
In particular, the proposed algorithm aims to keep the utility of the users that perform swap operation increasing for the first case, while focuses on the utility of channels for the second case. Additionally, both cases ensure an increasing sum utility of all players affected by the swap operation to achieve the stable matching.

\textbf{Case 1: $|\mathcal{U}^b|\!\le\!|\mathcal{M}|$.} First, we add $|\mathcal{M}|\!-\!|\mathcal{U}^b|$ virtual single-modal users to enable each channel to match a user, making the swap operation easier. Then, the user group set in the $b$-th cell is updated as ${\mathcal{G}_{\rm{A}}^b}\!=\!\mathcal{G}^b \cup \mathcal{U}_{\rm{0}}^b$, where $\mathcal{U}_{\rm{0}}^b$ is the established virtual user set with $|\mathcal{U}_{\rm{0}}^b|\!=\!|\mathcal{M}|\!-\!|\mathcal{U}^b|$. Correspondingly, we denote the updated index set of ${\mathcal{G}_{\rm{A}}^b}$ as ${\mathcal{Q}_{\rm{A}}^b}$. 

As the users belonging to a user group jointly decide the task performance, a bimodal user pair (\textit{rather than an individual bimodal user}) or a single-modal user is regarded as a selfish and rational player to make decisions, i.e., there are totally $|{\mathcal{G}_{\rm{A}}^b}|$ players in the $b$-th cell and each of them is denoted by $\mathcal{G}_q^b \in \mathcal{G}_{\rm{A}}^b$. Clearly, the resource sets of the players with the single-modal task and those with the bimodal task are different, which are $\mathcal{T}_{\rm{Si}}=\mathcal{T}$ and $\mathcal{T}_{\rm{Bi}}=\{(m,m',p,p'),\forall m,m' \in \mathcal{M}, m\ne m', \forall p,p' \in \mathcal{P}\}$, respectively. Thus, by denoting $\mathcal{T}_q^b$ as the resource set of player $\mathcal{G}_q^b$, we have $\mathcal{T}_q^b=\mathcal{T}_{\rm{Bi}}$ when $q\le N_{\rm{Bi}}^b$ and $\mathcal{T}_q^b=\mathcal{T}_{\rm{Si}}$, otherwise. 

Then, we can define a matching $\Phi$ as a function from set $\mathcal{G}_{\rm{A}} \cup \mathcal{T}_{\rm{A}}$ mapping into set $\mathcal{G}_{\rm{A}} \cup \mathcal{T}_{\rm{A}}$, where $\mathcal{G}_{\rm{A}}=\bigcup_{b\in{\mathcal{B}}}\mathcal{G}_{\rm{A}}^b$ and $\mathcal{T}_{\rm{A}}={\mathcal{T}_{\rm{Si}}} \cup {\mathcal{T}_{\rm{Bi}}}$. For a matching pair $(\mathcal{G}_q^b,t), t \in \mathcal{T}_{\rm{A}}$, we have $\Phi(\mathcal{G}_q^b)=t$ and $\Phi(t)=\mathcal{G}_q^b$. The utility function of player $\mathcal{G}_q^b$ under matching $\Phi$ is defined as
\begin{equation}
    U_q^b(\Phi)=\left\{
    \begin{split}
        &QoE_q^b,\quad {\rm{if}}\ q \in {\mathcal{Q}^b},\\
        &0, \quad \quad\quad {\rm{otherwise}}.
    \end{split}
    \right.
\end{equation}

Since a swap operation will affect not only the players in the current cell but also the players with the related channels in other cells, we define a coalition utility measuring the sum utility of all affected players as 
\begin{equation}
V(\Phi)=\sum\nolimits_{q \in {\mathcal{G}}_{\rm{af}}} U_q^b(\Phi),
\end{equation}
where ${\mathcal{G}}_{\rm{af}}$ is the player coalition formed by all players whose utility is affected by this swap matching.

Based on the above definitions, whether the considered matching is stable can be judged.

\begin{myDef}
A matching $\Phi$ is stable if and only if, for each player $\mathcal{G}_q^b \in \mathcal{G}_{\rm{A}}^b, \forall b \in \mathcal{B}$ with $\Phi(\mathcal{G}_q^b)=t$, there is no blocking matching $\Phi_{q}^{\rm{B}}$ such that, $\exists t'\in \mathcal{T}_q^b$:
\begin{enumerate}
    \item $\forall i \in \mathcal{G}_{(t)}^b\cup \mathcal{G}_{(t')}^b, U_i^b(\Phi_q^{(t,t')}) \ge U_i^b(\Phi_q^{\rm{B}})$ and
    \item $\exists i \in \mathcal{G}_{(t)}^b\cup \mathcal{G}_{(t')}^b$, $U_i^b(\Phi_q^{(t,t')}) > U_i^b(\Phi_q^{\rm{B}})$ and $V(\Phi_q^{(t,t')})>V(\Phi_q^{\rm{B}})$,
\end{enumerate}
where $\mathcal{G}_{(t)}^b$ is formed by the players in the $b$-th cell who have at least one same channel as those included in $t$, and $\Phi_q^{(t,t')}$ represents the swap matching, where the player $\mathcal{G}_q^b$ swap $t$ with $t'$, the players who are in $\mathcal{G}_{(t')}^b$ swap their channels accordingly, and the remaining players keep unchanged.
\end{myDef}

\textbf{Case 2: $|\mathcal{U}^b|>|\mathcal{M}|$.} Evidently, if we follow the solution for \textbf{Case 1}, no channel will be tentatively matched with $|\mathcal{U}^b|-|\mathcal{M}|$ of all users even if the overall QoE will increase. To deal with this problem, we focus on the market of channels to achieve the stable matching. 
However, when a channel is matched with a user with the bimodal task, its utility depends not only on which user it matches but on which channel the other user in the same group matches. Hence, we propose the user groups to perform the swap operation but the channels to make the swap decision based on their utility changes.

Similar to \textbf{Case 1}, we first add $|\mathcal{U}^b|-|\mathcal{M}|$ virtual channels to enable the swap operation at each user. Then, the channel set in the $b$-th cell can be updated as ${\mathcal{M}_{\rm{A}}^b}=\mathcal{M}\cup \mathcal{M}_{\rm{0}}^b$, where $\mathcal{M}_{\rm{0}}^b$ is the established virtual channel set with $|\mathcal{M}_{\rm{0}}^b|=|\mathcal{U}^b|-|\mathcal{M}|$. Then, the resource sets can be reformed in the similar manner but based on ${\mathcal{M}_{\rm{A}}^b}$. For simplicity, we use the same notation $\mathcal{T}_q^b$ to denote the resource set of each player $\mathcal{G}_q^b \in \mathcal{G}^b$. Furthermore, the utility function of channel $m\in {\mathcal{M}_{\rm{A}}^b}$ in the $b$-the cell under a matching $\Phi$ is defined as
\begin{equation}
   { U_m^b}'(\Phi)=\left\{
    \begin{split}
        &0, \quad \quad\quad\quad\  {\rm{if}}\  m \in \mathcal{M}_{\rm{0}}^b,\\
        &U_{\mu(m)}^b(\Phi)/2, {\rm{if}}\ m \in \mathcal{M}^b, \mu(m)\le N_{\rm{Bi}}^b,\\
        &U_{\mu(m)}^b(\Phi),\quad{\rm{if}}\ m \in \mathcal{M}^b,  \mu(m)> N_{\rm{Bi}}^b,\\
    \end{split}
    \right.
\end{equation}
where $\mu(m)\in\mathcal{Q}^b$ represents the index of the player that matches channel $m$. When the player is a bimodal user pair, ${ U_m^b}'(\Phi)$ is set as the half of the utility of the player. When the player is a single-modal user, ${ U_m^b}'(\Phi)$ is equal to the utility of the player. Thus, we give the following definition.
\begin{myDef}
A matching $\Phi$ is stable if and only if, for each channel $m \in {\mathcal{M}_{\rm{A}}^b}$ with $\mu(m)\in\mathcal{Q}^b$, there is no blocking matching $\Phi_{\mu(m)}^{\rm{B}}$ such that, $\exists t'\in \mathcal{T}_{\mu(m)}^b$:
\begin{enumerate}
    \item $\forall i \in \mathcal{M}_{({\mathcal{G}}^b)}, {U_i^b}'(\Phi_{\mu(m)}^{(t,t')}) \ge {U_i^b}'(\Phi_{\mu(m)}^{\rm{B}})$ and
    \item $\exists i \in \mathcal{M}_{({\mathcal{G}}^b)}$, ${U_i^b}'(\Phi_{\mu(m)}^{(t,t')}) > {U_i^b}'(\Phi_{\mu(m)}^{\rm{B}})$ and $V(\Phi_{\mu(m)}^{(t,t')})>V(\Phi_{\mu(m)}^{\rm{B}})$,
\end{enumerate}
where $\mathcal{M}_{({\mathcal{G}}^b)}$ is formed by all channels of all players in ${\mathcal{G}}^b, {\mathcal{G}}^b=\mathcal{G}_{(t)}^b\cup \mathcal{G}_{(t')}^b$.
\end{myDef}

Based on the above two definitions, we develop an efficient matching algorithm to obtain the stable matching, as shown in Algorithm 1. The proposed algorithm starts by initialing the matching channels and power levels of all users with a permutation of all channels and the minimum power level, respectively. Note that initialing the power level as the minimum one is to consume as less power as possible while achieving the same QoE. Then the users will search their resource set to find the blocking matching and thus update the current matching. Once no blocking matching is found, the stable matching $\Phi^*$ will be the output.
\setlength{\textfloatsep}{3pt}
\begin{algorithm}
\SetAlgoVlined 
\caption{ Proposed Matching Algorithm for Channel Assignment and Power Allocation}
\label{alg1}
\KwIn{$\mathcal{B},\ {\mathcal{U}^b}, \ {\mathcal{M}_{\rm{A}}^b},\ {\mathcal{G}_{\rm{A}}^b}, \ \mathcal{T}_q^b,\ \forall q\in \mathcal{Q}_{\rm{A}}^b, \ \forall b\in\mathcal{B}$.}
{\bf Initialization:} For each cell $b\in \mathcal{B}$, initial the matching channels and power levels of all users with a permutation of all channels and the minimum power level, respectively. Denote the current matching as $\Phi$.\\
\Repeat{No blocking matching is found}{
\For{all $b\in \mathcal{B}$}
{
\For{all $\mathcal{G}_q^b\in \mathcal{G}_{\rm{A}}^b$}
{
$t=\Phi(\mathcal{G}_q^b)$;\\
\For{all $t'\in \mathcal{T}_q^b,\ t'\ne t$}{
\eIf{$|\mathcal{U}^b|\le|\mathcal{M}|$}
{If $\Phi$ is a blocking matching according to Definition 2, update the matching as $\Phi=\Phi_q^{(t,t')}$; otherwise, keep the current matching state.}
{If $\Phi$ is a blocking matching according to Definition 3, update the matching as $\Phi=\Phi_q^{(t,t')}$; otherwise, keep the current matching state.
}
}}
}
}
\KwOut{The stable matching $\Phi^*$.}
\end{algorithm}

\vspace{-5pt}
\subsection{Analysis of the Proposed Matching Algorithm}
In this part, the proposed matching algorithm is analyzed from the aspects of convergence, stability, and complexity.

\textit{Convergence.} First, the number of transmitted semantic symbols, channels, user groups, and power levels are fixed, implying that the number of potential matching is finite. Second, the utility of user groups or channels is bounded by 1 and will increase monotonically by the swap operation. Therefore, the algorithm will terminate to a final matching after a finite number of iterations.

\textit{Stability.} The stability of the matching algorithm can be proved by the contradiction method. Taking \textbf{Case 1} as an example, we assume that, for player $\mathcal{G}_q^b$, a matching $\Phi_j$ is reached with $\Phi_j(\mathcal{G}_q^b)=t$ when the algorithm is converged and it satisfies: $\exists t'\in \mathcal{T}_q^b$ such that
\begin{enumerate}
    \item $\forall i \in \mathcal{G}_{(t)}^b\cup \mathcal{G}_{(t')}^b, U_i^b(\Phi_q^{(t,t')}) \ge U_i^b(\Phi_j)$ and
    \item $\exists i \in \mathcal{G}_{(t)}^b\cup \mathcal{G}_{(t')}^b$, $U_i^b(\Phi_q^{(t,t')}) > U_i^b(\Phi_j)$ and $V(\Phi_q^{(t,t')})>V(\Phi_j)$.
\end{enumerate}
Clearly, $\Phi_j$ is a blocking matching, and it will be updated by $\Phi_q^{(t,t')}$ according to Algorithm 1, which indicates that the algorithm will not stop at $\Phi_j$. Evidently, there is a contradiction with the initial assumption. Hence, the proposed algorithm will converge to a stable matching.

\textit{Complexity.} In each iteration, each user group attempts to find the blocking matching through searching its resource set. Considering the worst case, the maximum numbers of searching operations for single-modal users and bimodal user groups are $|{\mathcal{T}_{\rm{Si}}}|=|\mathcal{M}_{\rm{A}}^b|\times|\mathcal{P}|$ and $ |{\mathcal{T}_{\rm{Bi}}}|=\tbinom{2}{|\mathcal{M}_{\rm{A}}^b|}\times|\mathcal{P}|^2$, respectively. Considering the number of cells and the number of users, we can obtain the worst-case complexity of Algorithm 1 as $\mathcal{O}(I_{\rm{ite}} \times |\mathcal{B}|\times((|\mathcal{G}_{\rm{A}}^b|-N_{\rm{Bi}}^b)\times|\mathcal{M}_{\rm{A}}^b|\times|\mathcal{P}|+N_{\rm{Bi}}^b \times \tbinom{2}{|\mathcal{M}_{\rm{A}}^b|}\times|\mathcal{P}|^2))$, where $I_{\rm{ite}}$ is the number of iterations. 

In contrast, the complexity of the exhaustive search here is as $\mathcal{O}(({\frac{{\
\rm{max}}(|\mathcal{U}^b|,M)!}{{\
\rm{min}}(|\mathcal{U}^b|,M)!}\times |\mathcal{P}|^{|\mathcal{U}^b|}})^{|\mathcal{B}|})$, which grows exponentially with the number of cells and users. Apparently, the proposed algorithm yields much lower computation complexity although returning a suboptimal solution.

\vspace{-5pt}
\section{Simulation results}
In this section, we first present the simulation results of semantic fidelity curve fitting, and then evaluate the performance of the proposed QoE-aware resource allocation method in a multi-cell multi-task network. 

\subsection{Semantic Fidelity Curve Fitting}
Two DNNs are trained to approximate the semantic fidelity of the considered single-modal and bimodal tasks, respectively. To build the training dataset, DeepSC and DeepSC-VQA are exploited to obtain the two discrete mapping tables, respectively. They are $\xi_{q}^b=f_{\rm{Si}}(k_{u_{\rm{s}}},\gamma_{u_{\rm{s}}}^b)$ and $\xi_{q}^b=f_{\rm{Bi}}(k_{u_{\rm{bt}}}, k_{u_{\rm{bi}}},\gamma_{u_{\rm{bt}}}^b,\gamma_{u_{\rm{bi}}}^b)$, where $k_{u_{\rm{s}}} \in \{1,2,\dots,20\}\ \rm{symbols/word}$, $k_{u_{\rm{bt}}}\in \{1,2,4,6,8,10,12\}\ \rm{symbols/word}$, $k_{u_{\rm{bi}}}\in \{1,2,4,6,8,12,16\}*197\ \rm{symbols/image}$, and $\gamma_{u_{\rm{s}}}^b, \gamma_{u_{\rm{bt}}}^b,\gamma_{u_{\rm{bi}}}^b \in \{-10,$ $-9,\dots,20\}\ \rm{dB}$.
The DNN for the single-modal task is made up of an input layer, a fully connected hidden layers, and an output layer, containing 2, 16, and 1 neurons, respectively. Similarly, the DNN for the bimodal task is made up of an input layer, two fully connected hidden layers, and an output layer, containing 4, 32, 32, and 1 neurons, respectively. The sigmoid function is used as the activation function since the semantic fidelity is between 0 and 1. The Adam optimizer is adopted with the learning rate of 0.001.

\begin{figure}
\vspace{-10pt}
  \centering
  \includegraphics[width=0.38\textwidth]{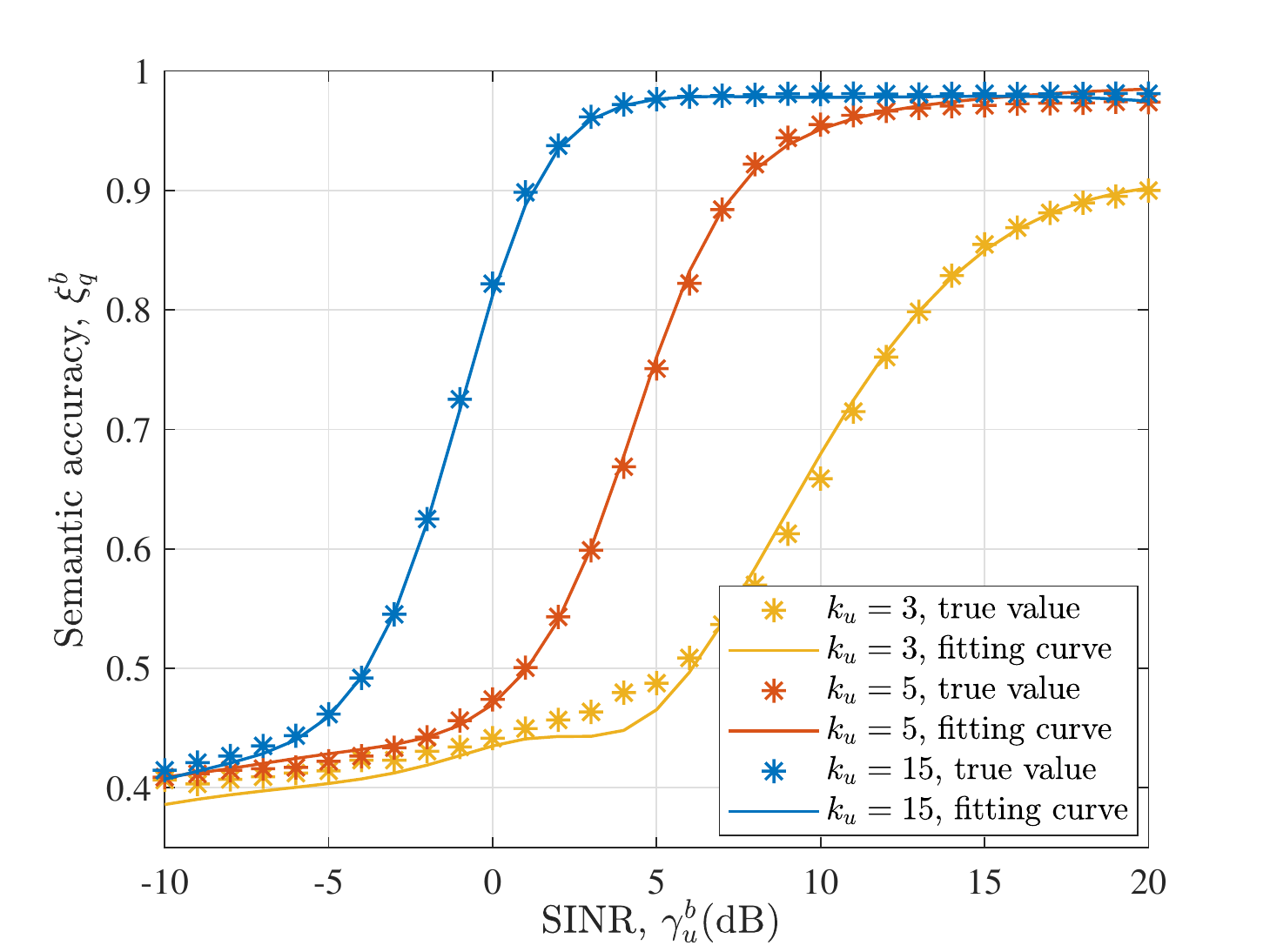}
  \caption{The fitting performance of the single-modal task.}
  \vspace{-3pt}
  \label{fitting_Si}
\end{figure}

\begin{figure}
\vspace{-10pt}
  \centering
  \includegraphics[width=0.38\textwidth]{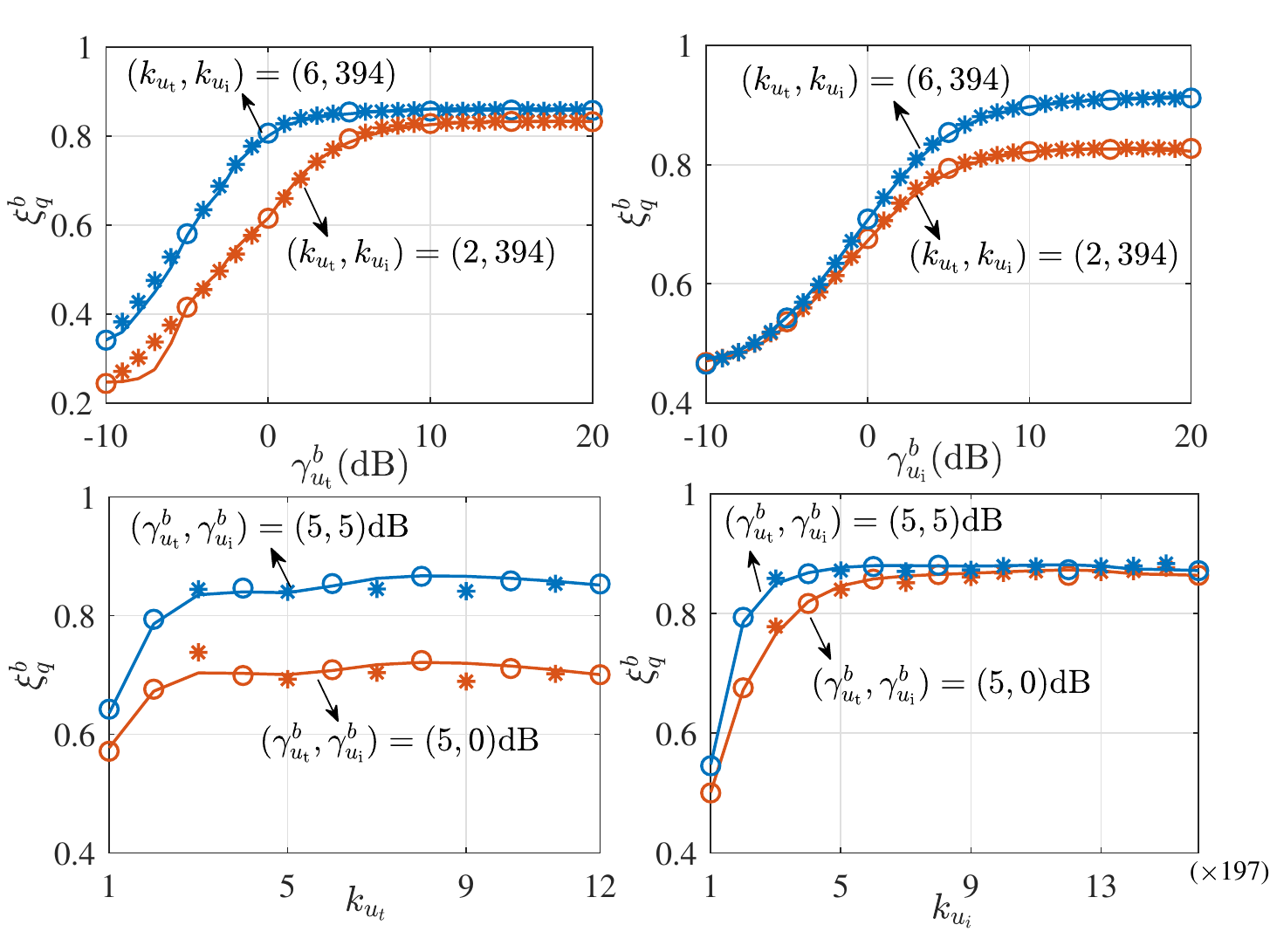}
  \caption{The fitting performance of the bimodal task ($\circ$: True value for training; *: True value for verification; --: Fitting curve). Default parameters: $\gamma_{u_{\rm{bt}}}^b=\gamma_{u_{\rm{bi}}}^b=5\ \rm{dB}$, $k_{u_{\rm{bt}}}=2\ \rm{symbols/word}$, and $k_{u_{\rm{bi}}}=394\ \rm{symbols/image}$.}
  \vspace{-3pt}
  \label{fitting_Bi}
\end{figure}

The fitting performance of the developed DNNs for the considered two tasks is evaluated in Fig.~\ref{fitting_Si} and Fig.~\ref{fitting_Bi}. From Fig.~\ref{fitting_Si}, the semantic fidelity of the single-modal task is well approximated by the developed DNN. As for the bimodal task, we present $\xi_q^b$ versus $\gamma_{u_{\rm{bt}}}^b$, $\gamma_{u_{\rm{bi}}}^b$, $k_{u_{\rm{bt}}}$, and $k_{u_{\rm{bi}}}$ in Fig.~\ref{fitting_Bi}. The neural network has a good fitting for the first two cases while performs a little worse for the others. This is because different network models are trained for different $(k_{u_{\rm{bt}}}, k_{u_{\rm{bi}}})$ in the latter two cases, resulting in small performance fluctuations. Additionally, the fitting performance of the DNN designed for the bimodal task is verified by more data that are not in the train dataset as well. 

\vspace{-5pt}
\subsection{Performance of the Proposed Method}
We consider three cells in the simulation, where users are randomly deployed. The radius of each cell is set as $500\ {\rm{m}}$. Each BS is equipped with two receiving antennas and each user is with single antenna. For the channel model, both large-scale fading and small-scale Rayleigh fading are considered. The adopted pathloss model is $128.1+37.6\rm{lg[d(km)]\  dB}$ and the shadowing factor is set as $6\ {\rm{dB}}$. The bandwith of each channel is $W=180\ \rm{kHz}$. The noise power spectral density is $\sigma^2=-174\ {\rm{dBm/Hz}}$. 
The available power level set is $\mathcal{P}=\{-10, -5, 0, 5, 10, 15, 20\}\ {\rm{dBm}}$. 
Meanwhile, each user generates its QoE-related parameters as $w_u\sim U(0,1)$, $\xi_u^{\rm{req}}\sim U(0.8,0.9)$, and $\lambda_u \sim N(55,2.5^2)$. In addition, for text transmission users, we set $\varphi_u^{\rm{req}}\sim U(50,70)$ in Ksuts/s and $\beta_u \sim N(0.2, 0.05^2)$. For image transmission users, we set $\varphi_u^{\rm{req}}\sim U(80,100)$ in Ksuts/s and $\beta_u \sim N(0.1, 0.02^2)$. Unless specifically stated, we set $(N_{\rm{Si}}, N_{\rm{Bi}})=(6,6)$, $|\mathcal{M}|=6$, and $G_{\rm{th}}=0.5$. 

The DQNs for the semantic compression subproblem of the single-modal task and the bimodal task are as follows. The model for the single-modal task is made up of an input layer ($|s_t|$ neurons, $s_t \in {\mathcal{S}}_{\rm{Si}}$), three fully connected hidden layers (256, 256, and 256 neurons), and an output layer ($|{\mathcal{A}}_{\rm{Si}}|$ neurons). The model for the bimodal task consists of an input layer ($|s_t|$ neurons, $s_t \in {\mathcal{S}}_{\rm{Bi}}$), three fully connected hidden layers (256, 256, and 256 neurons), and an output layer ($|{\mathcal{A}}_{\rm{Bi}}|$ neurons). The learning rates of the two models are set to $10^{-4}$ and $5 \times 10^{-5}$, respectively. For both models, the activation function is the relu function and the Adam optimizer is adopted. The explore rate, $\epsilon$, is linearly annealed from 1 to 0.02 over the beginning 1800 episodes and remains constant over the last 200 episodes.

We first verify the superiority of the developed QoE based formulation. Fig.~5 compares the QoE maximization and S-R maximization methods\cite{OurWork}. Here, the upper bound is obtained by assuming that the maximum number of users in each cell, i.e., ${\rm{max}}(|\mathcal{U}^b|,|\mathcal{M}|)$, can be served and their QoE can reach 1. Thus the upper bound is irrelevant to $G_{\rm{th}}$, which is compared to demonstrate the effectiveness of the proposed method. As $G_{\rm{th}}$ increases, the overall QoE of the S-R maximization method decreases while that of the proposed method keep almost stable, which implies that   the proposed method can better adapt to the changes in the user requirements.

\begin{figure}
\vspace{-10pt}
  \centering
  \includegraphics[width=0.38\textwidth]{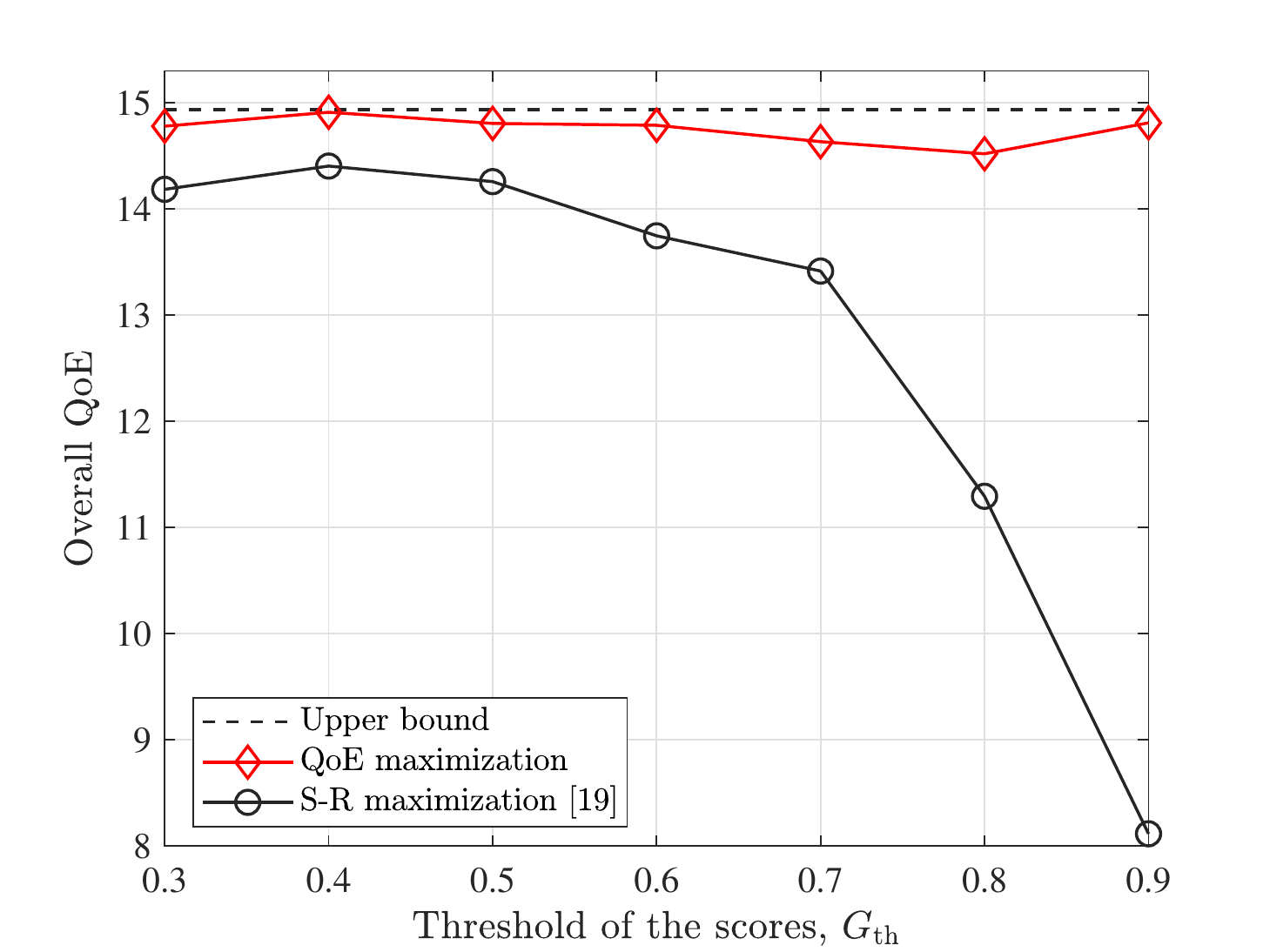}
  \caption{Comparison with the benchmark \cite{OurWork}.}
  \vspace{-3pt}
\end{figure}

Fig.~6 evaluates the performance of the proposed algorithm. Due to the unacceptable computation complexity of the exhaustive search method as analyzed in Section IV-D, we do not compare the proposed matching algorithm with that but with the upper bound of overall QoE in Fig.~6(a). From this figure, the proposed matching algorithm outperforms the random matching method significantly and is very close to the upper bound. In addition, compared with the algorithm in \cite{OurGlobecom}, the proposed algorithm can improve the overall QoE effectively. This is because it considers the coalition utility in Definitions 2 and 3, and the task performance is better characterized by curve fitting. The convergence of the proposed algorithm is validated in Fig.~6(b). As the number of swap operations increases, the overall QoE increases gradually and finally reaches the maximum value. Furthermore, the number of swap operations required for a stable matching increases with $|\mathcal{M}|$.  
\begin{figure}
\vspace{-15pt}
	\centering
	\subfigcapskip=-5pt
	\subfigure[Comparisons with different algorithms.]{
		\includegraphics[width=0.76\linewidth]{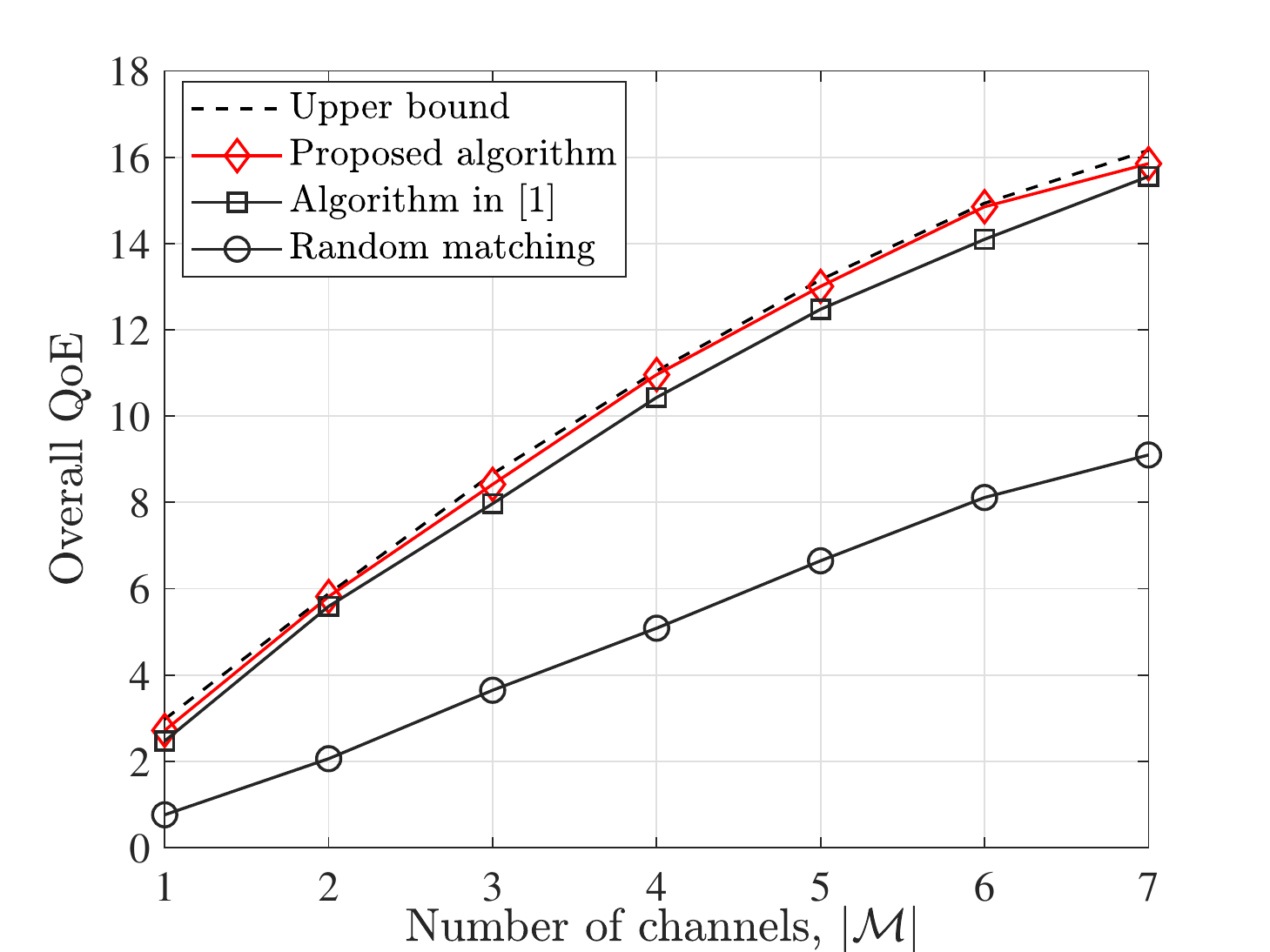}}
	\vspace{-5pt}
	\subfigure[The overall QoE versus the number of swap operations.]{
		\includegraphics[width=0.76\linewidth]{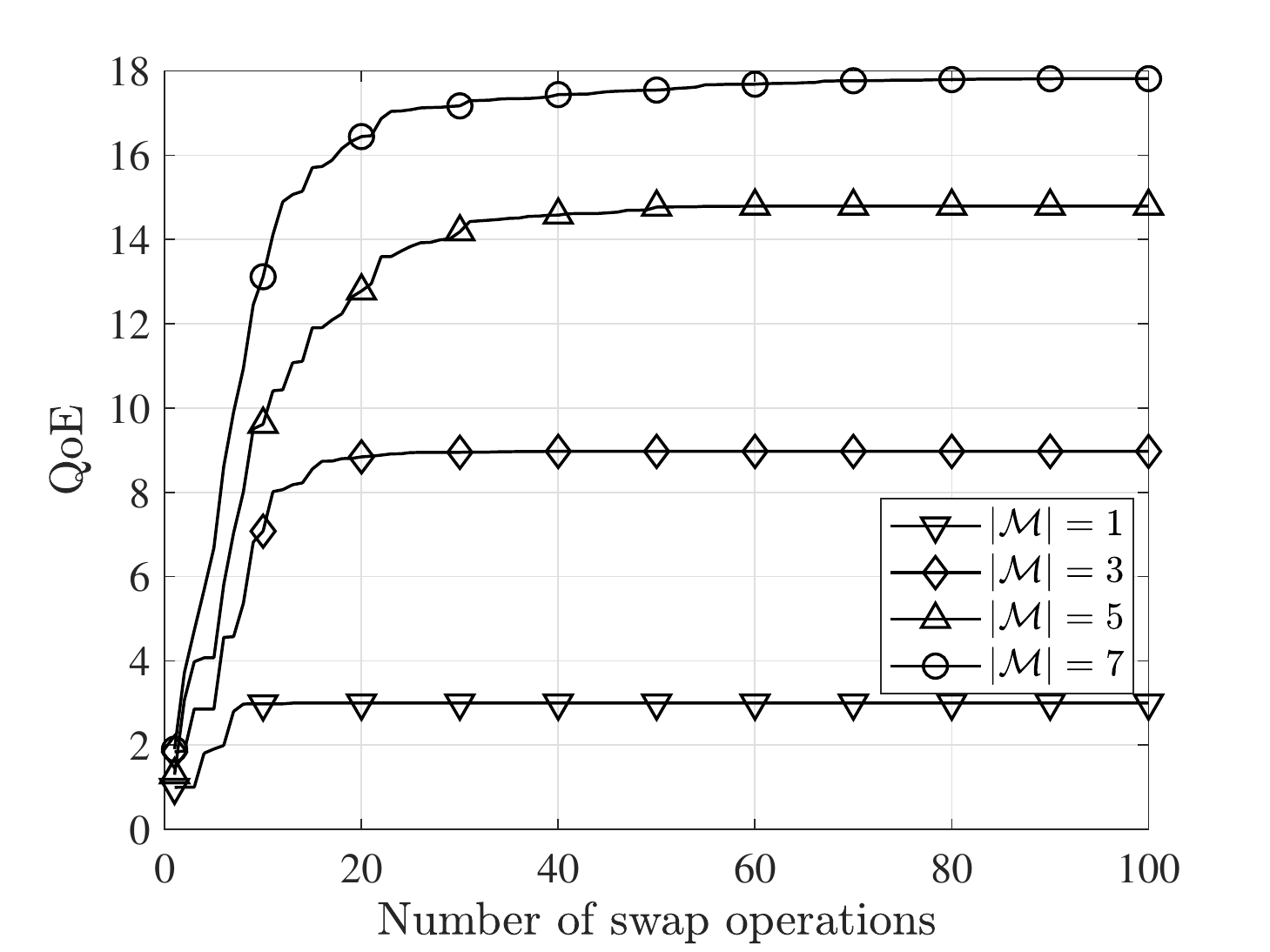}}
	\caption{The performance of the proposed algorithm.}
	\vspace{-3pt}
\end{figure}

Fig.~7 compares the proposed semantic-aware resource allocation model with the conventional ones. The channel assignment and power allocation results of the conventional models are obtained by maximizing the sum bit rate, and then the overall QoE is solved with different semantic compression ratio. Here, ``Conventional model 1'', ``Conventional model 2'', ``Conventional model 3'', and ``Conventional model 4'' represent the conventional model with $k_{u_{\rm{s}}}\!=\!k_{u_{\rm{bt}}}\!=\!1\ \rm{symbols/word}$ and $k_{u_{\rm{bi}}}\!=\!1\! \times\! 197\ \rm{symbols/image}$, $k_{u_{\rm{s}}}\!=\!k_{u_{\rm{bt}}}\!=\!3\ \rm{symbols/word}$ and $k_{u_{\rm{bi}}}\!=\!3\! \times\! 197\ \rm{symbols/image}$, $k_{u_{\rm{s}}}\!=\!k_{u_{\rm{bt}}}\!=\!5\ \rm{symbols/word}$ and $k_{u_{\rm{bi}}}\!=\!5\! \times\! 197\ \rm{symbols/image}$, and $k_{u_{\rm{s}}}\!=\!k_{u_{\rm{bt}}}=\!7\ \rm{symbols/word}$ and $k_{u_{\rm{bi}}}\!=\!7\! \times 197\ \rm{symbols/image}$, respectively. Additionally, the optimal $k_u, u \in \bigcup_{b\in{\mathcal{B}}}\mathcal{U}^b$ is obtained by optimizing the QoE of each user based on the channel assignment and power allocation results. From Fig. 7(a), the proposed model greatly outperforms the conventional one with different semantic compression, even with the optimal $k_u$. Moreover, the overall QoE of ``Conventional model 4'' first decreases and then increases when $|\mathcal{M}|=2$. This is because if the two channels are assigned to a bimodal user pair, the overall QoE could be 0 as the user requirement of image transmission users cannot be satisfied. In addition, Fig. 7(b) shows that the overall QoE of the proposed model increases with the number of users, and is larger than that of the conventional model with the optimal $k_u$.
\begin{figure}
\vspace{-10pt}
	\centering
	\subfigure[The overall QoE versus the number of channels.]{
		\includegraphics[width=0.76\linewidth]{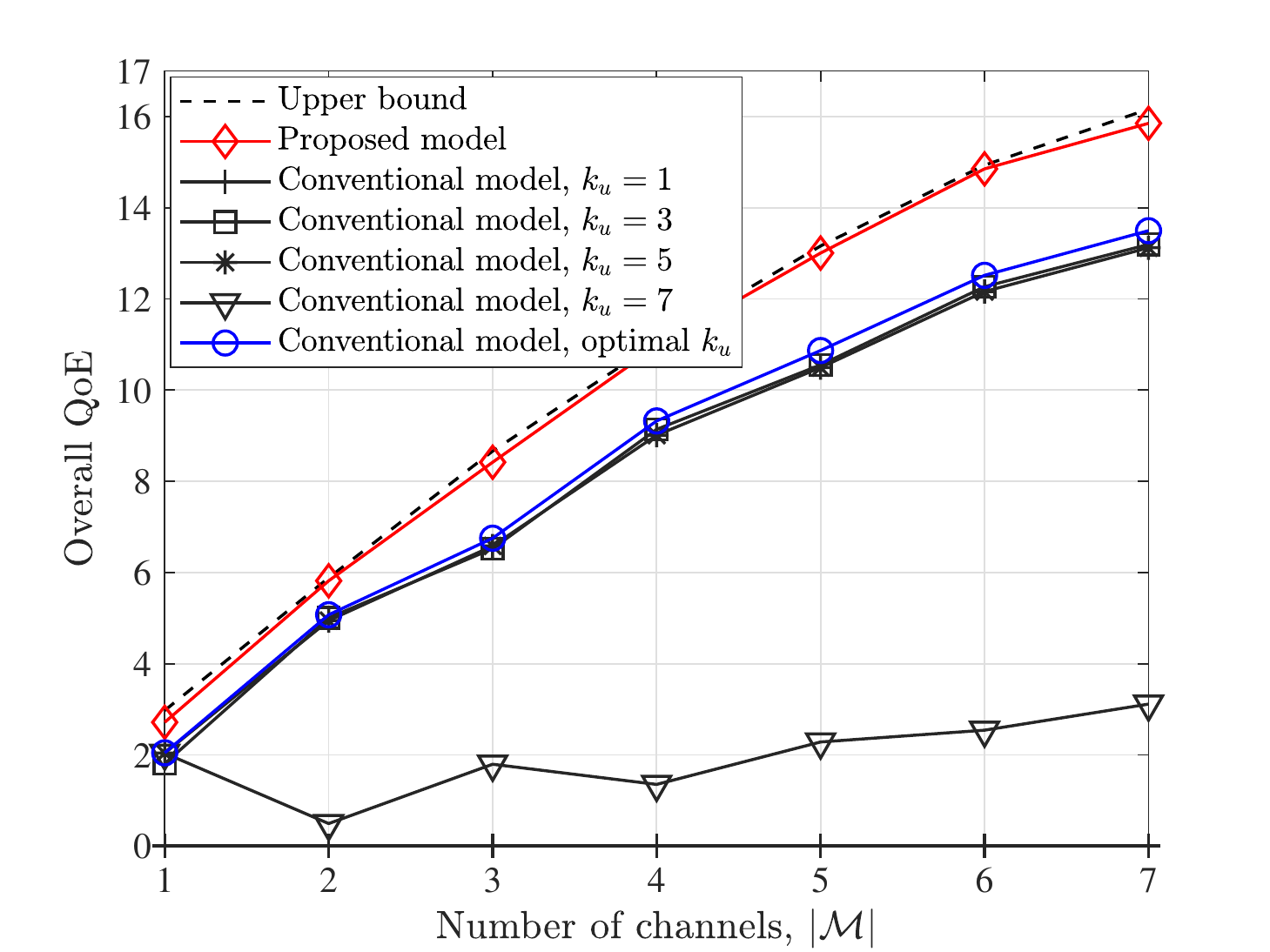}}
	\subfigure[The overall QoE versus the number of users.]{
		\includegraphics[width=0.76\linewidth]{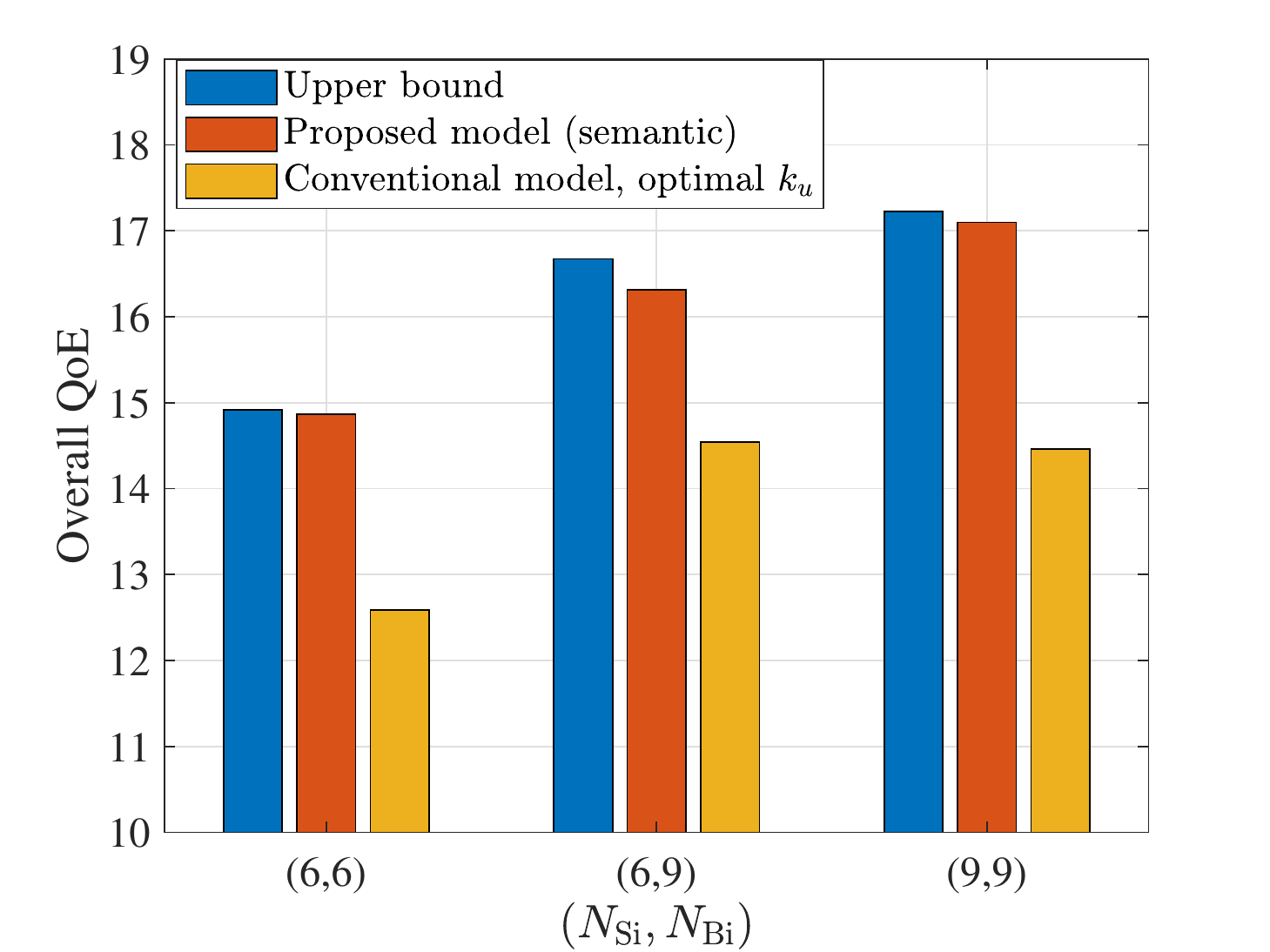}}
	\caption{Comparisons with conventional models.}
	\vspace{-3pt}
\end{figure}

Fig.~8 evaluates the proposed method with or without multi-cell cooperation. The method with multi-cell cooperation yields better performance than the one without multi-cell cooperation, which verifies the effectiveness of the proposed method in coping with the inter-cell interference.

\begin{figure}
\vspace{-10pt}
  \centering
  \includegraphics[width=0.38\textwidth]{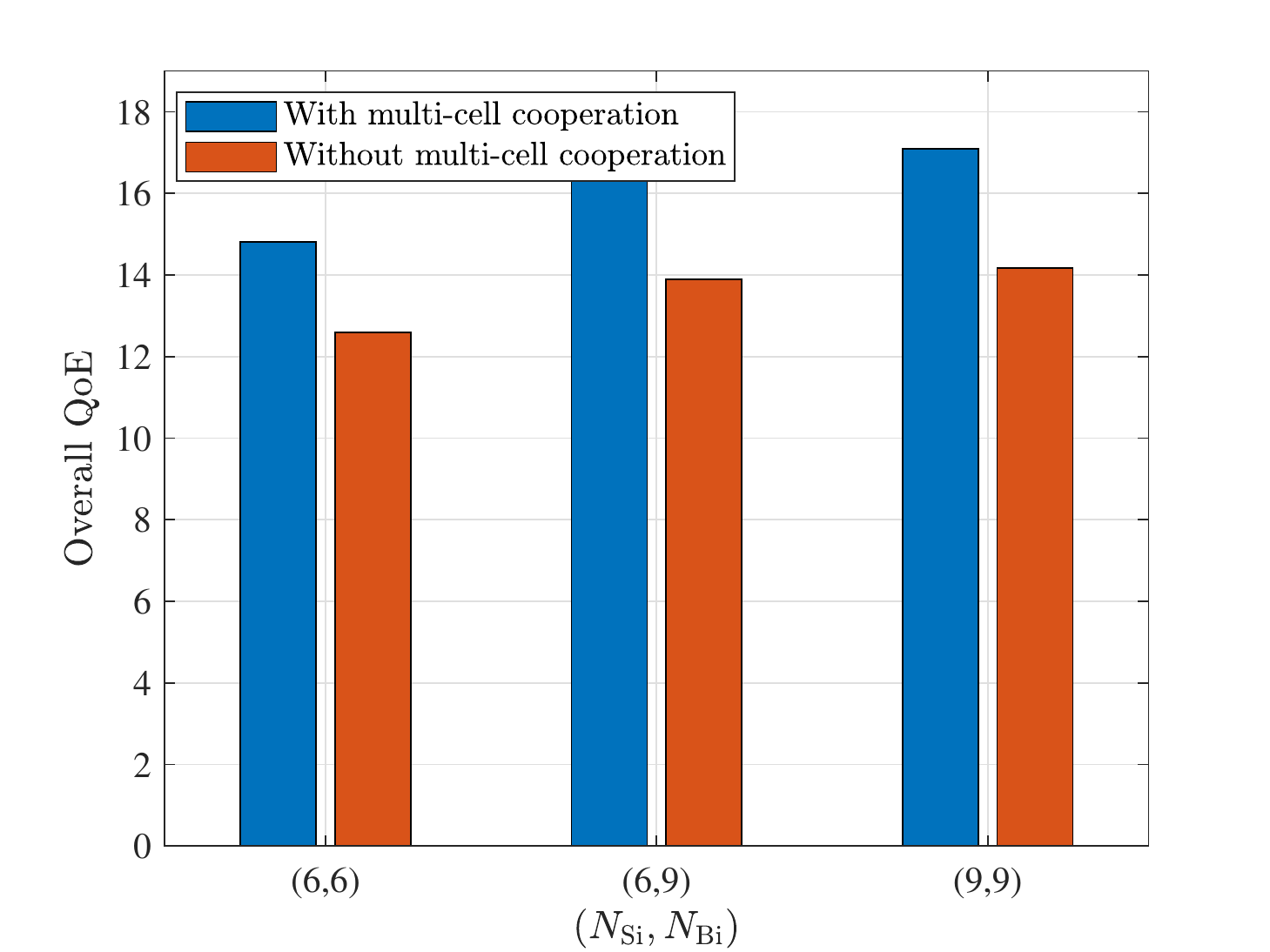}
  \caption{The proposed method with or without multi-cell cooperation.}
  \vspace{-3pt}
\end{figure}

Fig.~9 shows the overall QoE and power consumption of the proposed method under different settings, respectively. From Fig.~9(a), the overall QoE increases with the number of users in the network. However, only little performance degradation in the overall QoE can be seen when $N_{\rm{r}}$ and $|\mathcal{P}|$ decrease, which is because more transmit power will be allocated to compensate users for the QoE, as shown in Fig.~9(b). Besides, there is no significant difference of the power consumption between the cases with different $(N_{\rm{Si}},N_{\rm{Bi}})$. This is because the interference experienced in the case where $(N_{\rm{Si}},N_{\rm{Bi}})=(4,4)$ may be smaller than the other case and thus the allocated transmit power will be higher to improve the QoE.

\begin{figure}
\vspace{-10pt}
	\centering
	\subfigcapskip=-5pt
	\subfigure[The overall QoE.]{
		\includegraphics[width=0.76\linewidth]{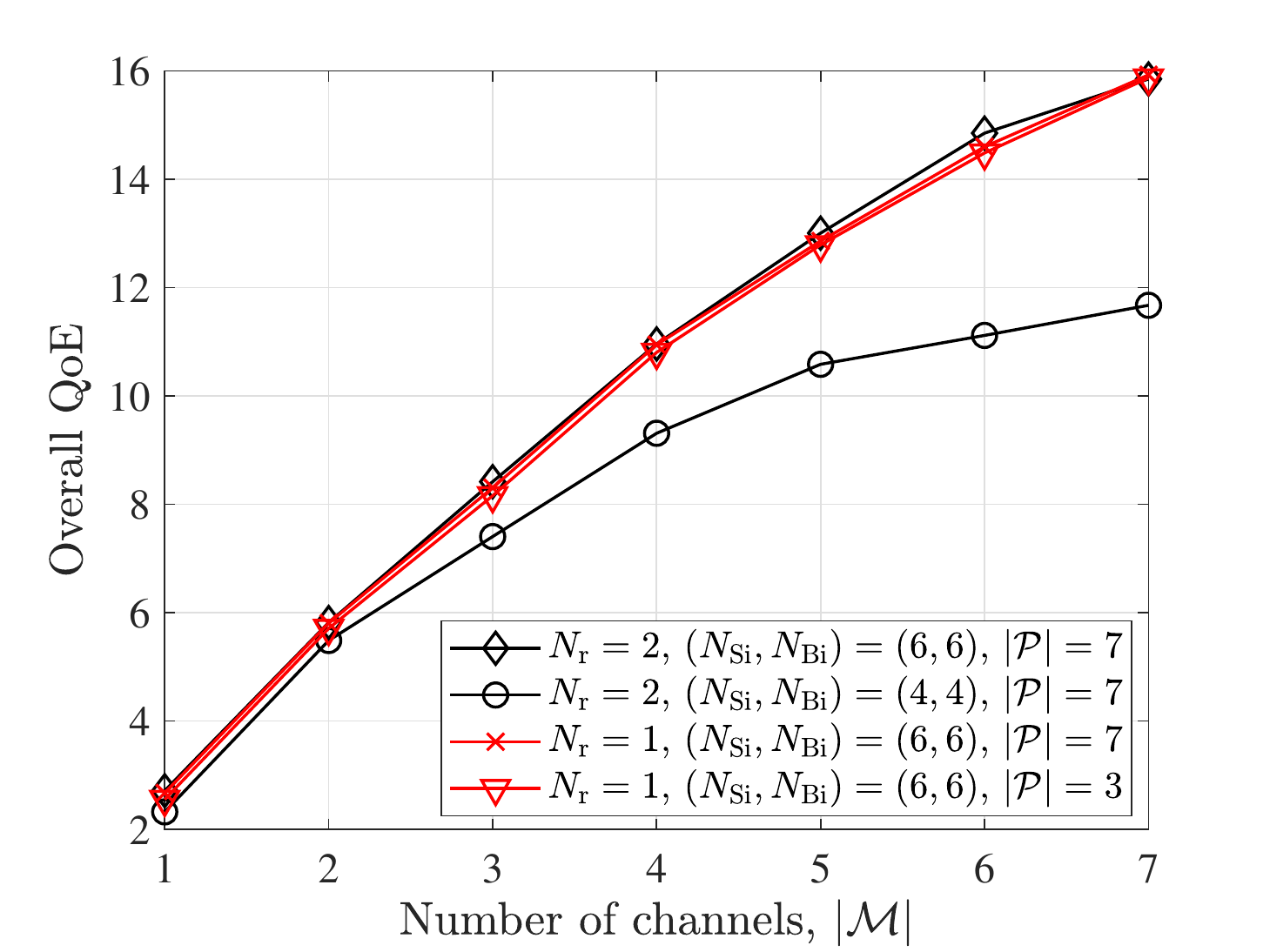}}
	\vspace{-5pt}
	\subfigure[The overall power consumption.]{
		\includegraphics[width=0.76\linewidth]{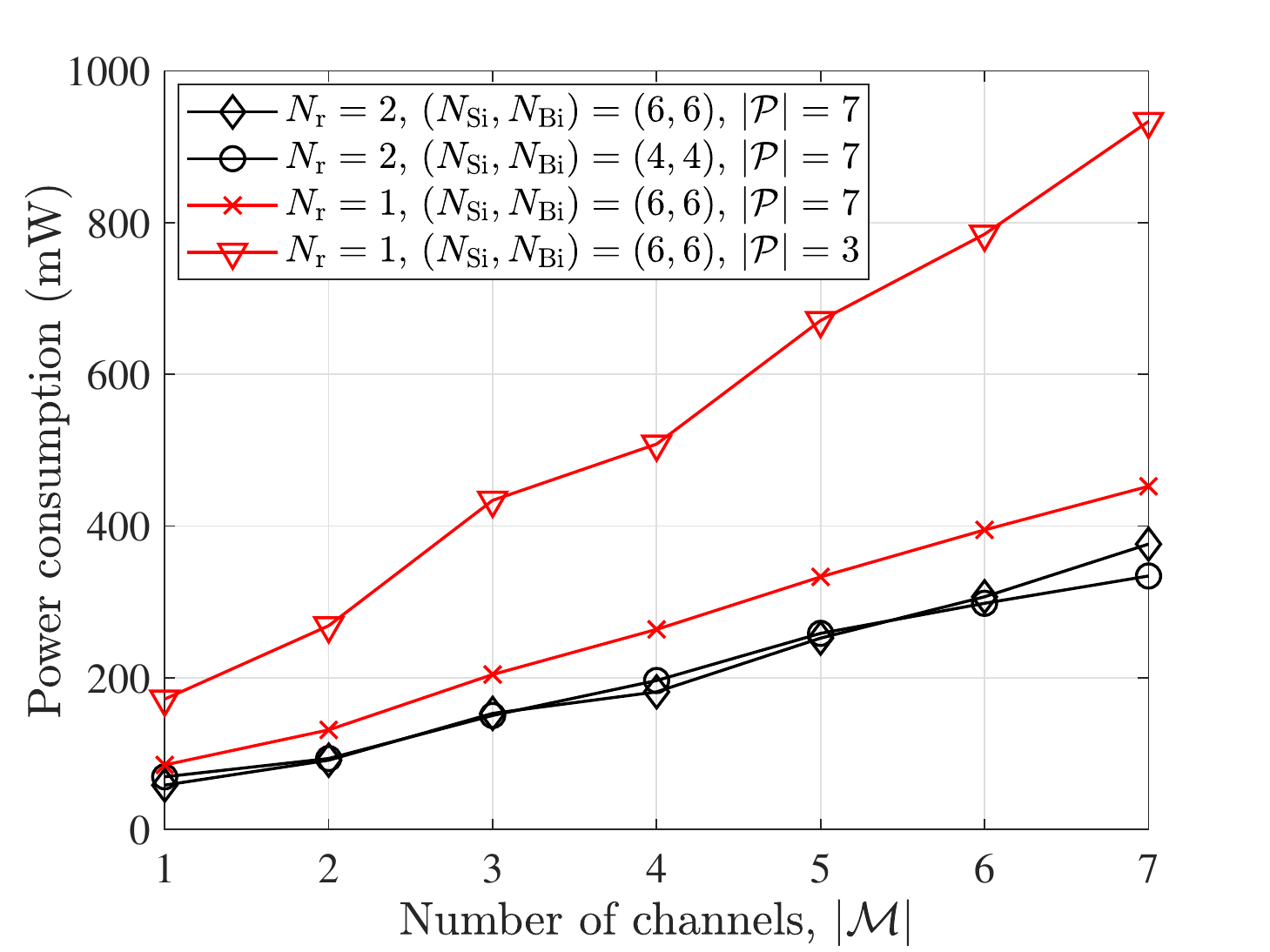}}
	\caption{The proposed method under different settings.}
	\vspace{-3pt}
\end{figure}

\vspace{-10pt}
\subsection{Compatibility Evaluation}
In this part, we evaluate the compatibility of the proposed semantic-aware resource allocation formulation from the QoE and QoS perspectives. The compatibility is reflected in the reasonable performance by fair resource allocation for all kinds of users when enabling equivalent QoS performance for all users. To verify whether the resources are allocated as we expected, we assume that the users with the conventional communication system perform the same tasks as those with the semantic one, i.e., the text transmission task and the VQA task, and these users with the same task should have the same communication requirements. In addition, we assume that $(N_{\rm{Si}},N_{\rm{Bi}})=(6,6)$, half of the users are semantic users, and the others are conventional users. For the QoE evaluation, the QoE-related parameter settings for the conventional users are as follow. We set ${C_u^{\rm{req}}}\sim U(0.4,0.6)$ in Mbps and ${\beta_u^{\rm{C}}} \sim N(30, 2^2)$ for the single-modal task. As for the bimodal task, we set ${C_u^{\rm{req}}}\sim U(0.8,1.2)$ in Mbps and ${\beta_u^{\rm{C}}} \sim N(15, 1^2)$ for text transmission users, and ${C_u^{\rm{req}}}\sim U(4.5,6.5)$ in Mbps and ${\beta_u^{\rm{C}}} \sim N(2, 0.1^2)$ for image transmission users. On the other hand, to evaluate the QoS performance in such a network, for the users with both systems, we set $\Gamma_u^{\rm{req}} \sim U(40,63)$ in Ksuts/s for text transmission users and $\Gamma_u^{\rm{req}} \sim U(64,94)$ in Ksuts/s for image transmission users. In addition, the transforming factors are set as $\mu_{\rm{text}}=40$ bits per word \cite{OurWork} for text transmission users and $\mu_{\rm{image}}=55624$ bits per image \cite{DeepSC-VQA} for image transmission users, respectively. To make the resource allocation for the users with both systems as fair as possible, we set these parameters for the conventional system based on Eq. (16), ensuring the settings for each task keep consistent with that for the semantic system.

Fig.~10 illustrates the overall QoE of the users in the network where semantic and conventional communications coexist. The upper bound for the sum QoE is the same as that in Fig.~6(a), and the upper bound of the QoE for the semantic system is equal to that for the conventional system as we set $(N_{\rm{Si}}^{\rm{Sem}},N_{\rm{Bi}}^{\rm{Sem}})=(N_{\rm{Si}}^{\rm{Con}},N_{\rm{Bi}}^{\rm{Con}})=(3,3)$ in the simulation. The QoE of each system increases with $|\mathcal{M}|$ since more users will be served with increasing available channels. In addition, we observe that (i) the gap between the sum QoE and the upper bound becomes wider as $|\mathcal{M}|$ increases, (ii) the QoE of the semantic system becomes closer to the upper bound, and (iii) the QoE of the conventional system is much smaller than that of the semantic system. The reason is that the bimodal users with the conventional system can hardly be served since it is difficult for them to achieve the transmission rate threshold, especially for image transmission users. As shown in Fig. 10(b), the number of served users with bimodal tasks is always zero, which indicates that the conventional system has no advantage in the tasks that focus on the effective execution at the receiver and the semantic system has much stronger compression ability. Thus, to improve the overall QoE of the network, more resources will be assigned to semantic users, facilitating the resource utilization maximization. Consequently, the QoE of the semantic system will approach the upper bound due to more available channels but limit users, and that of the conventional system will approach 3 as there are 3 single-modal users in total. Moreover, since there should be more bimodal users with the conventional systems served with increasing $|\mathcal{M}|$ while none of them achieves the requirement, the sum QoE becomes further away from the upper bound. In addition, as two users in a bimodal user pair joint decide the task performance, the bimodal user pairs are more competitive than the users with the single-modal task, especially when the number of available channels is only 2. Hence, the number of served users with the bimodal task is larger than that with the single-modal task when $|\mathcal{M}|=2$.
\begin{figure}
\vspace{-10pt}
	\centering
	\subfigcapskip=-5pt
	\subfigure[The overall QoE.]{
		\includegraphics[width=0.76\linewidth]{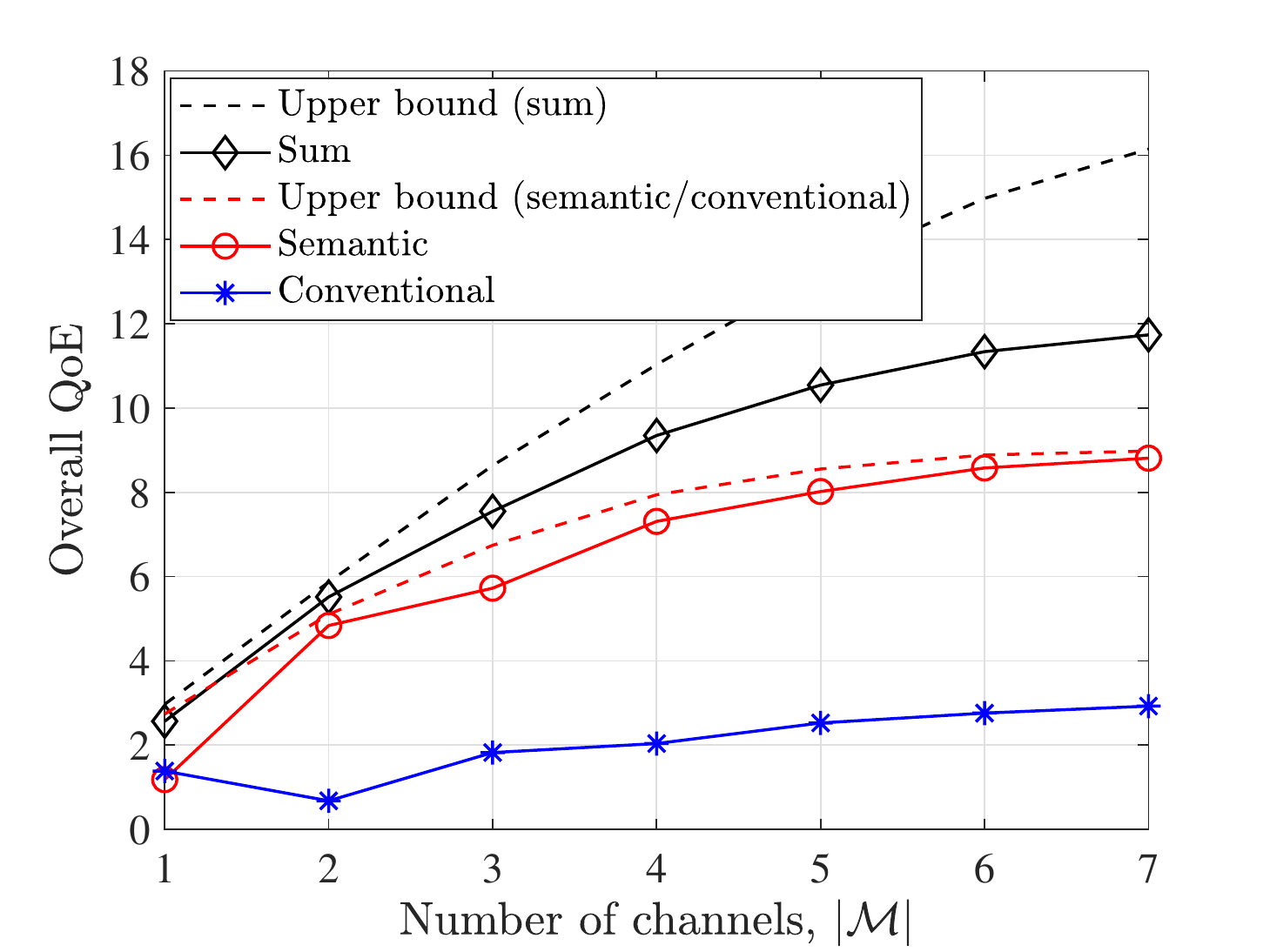}}
	\vspace{-5pt}
	\subfigure[The average number of served users in the network.]{
		\includegraphics[width=0.76\linewidth]{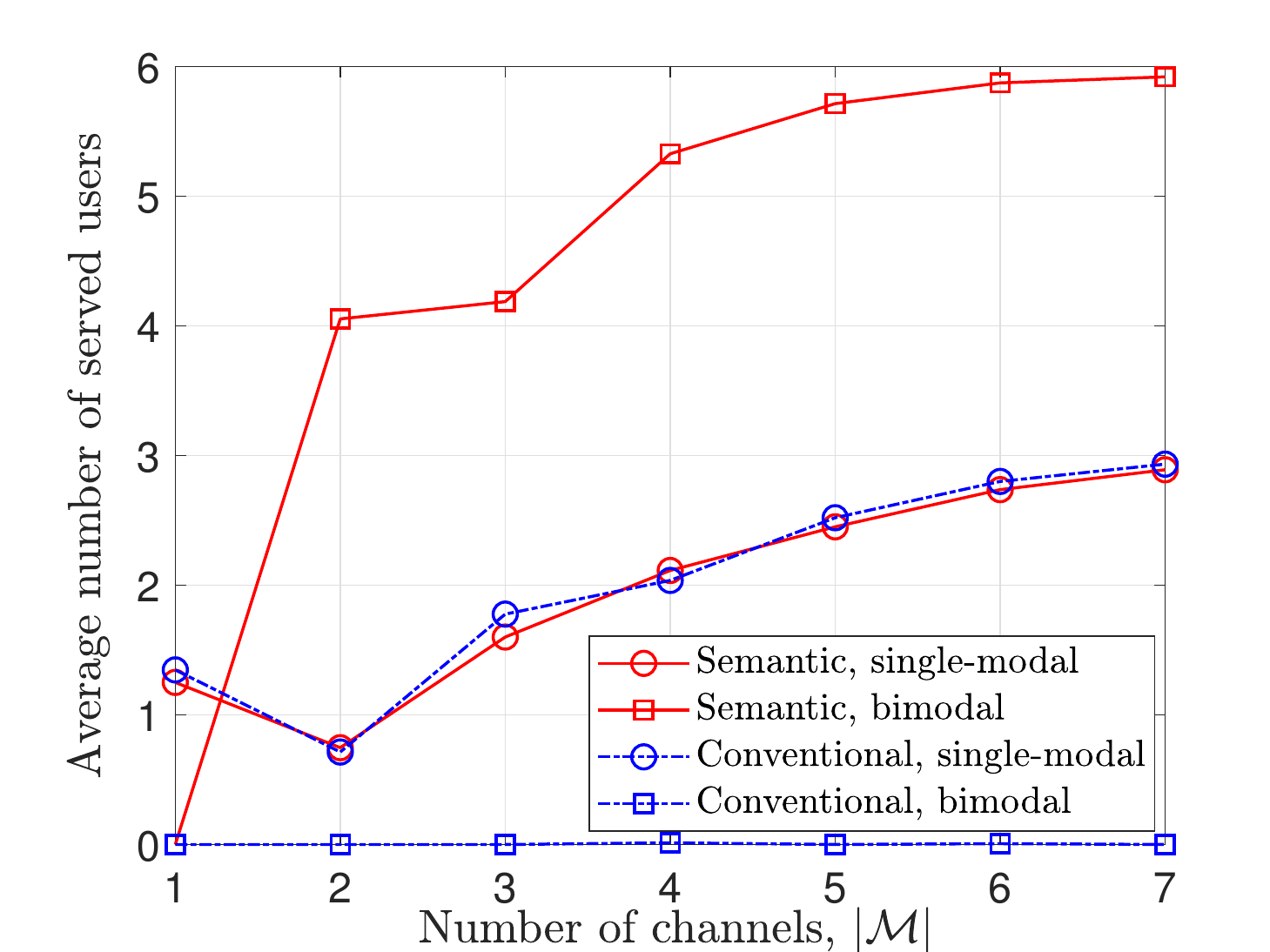}}
	\caption{The QoE performance.}
	\vspace{-3pt}
\end{figure}

Fig.~11 verifies the compatibility of the proposed resource allocation model in the QoS performance. Although the upper bound of S-R is hard to obtained, the similar trend can be observed as that in Fig.~10. 
\begin{figure}
\vspace{-10pt}
	\centering
	\subfigcapskip=-5pt
	\subfigure[The S-R.]{
		\includegraphics[width=0.76\linewidth]{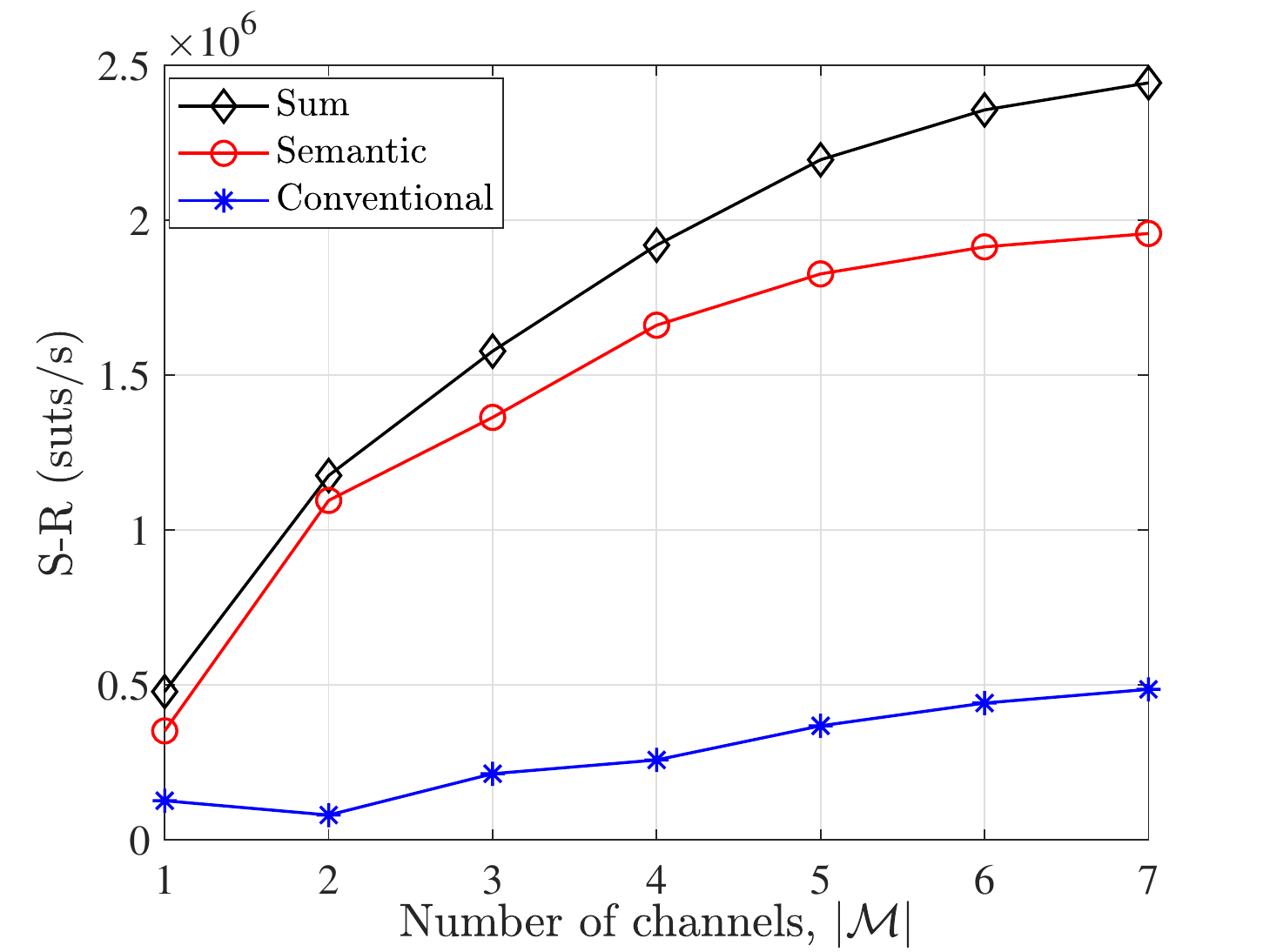}
		\vspace{-10pt}}
	\subfigure[The average number of served users in the network.]{
	\vspace{-10pt}
		\includegraphics[width=0.76\linewidth]{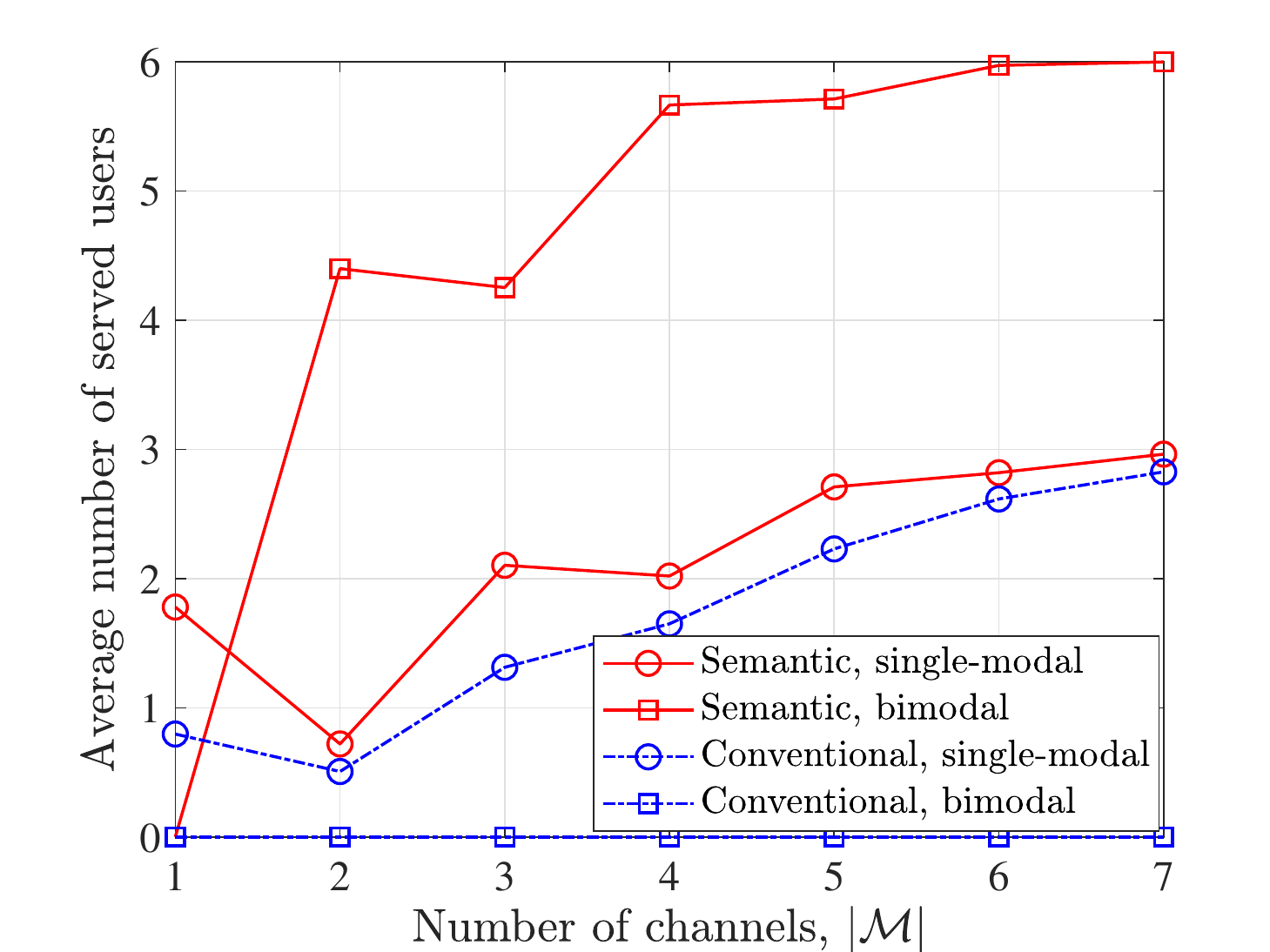}}
	\caption{The S-R performance.}
	\vspace{-5pt}
\end{figure}

\section{Conclusion}
In this paper, we studied the semantic-aware resource allocation in multi-task multi-cell networks. Specifically, semantic entropy was defined and quantified with an approximate measure, which made it possible to deal with the coexistence of multiple tasks for the semantic-aware resource allocation. Subsequently, we developed a novel QoE model in terms of semantic rate and semantic fidelity to better reflect the user experience. In addition, we formulated a QoE based semantic-aware resource allocation model, which showed excellent compatibility with conventional communications. Finally, we proposed a solution that combines deep Q-network and matching theory. 
Simulation results illustrated that the developed formulation can provide higher user satisfaction than the semantic transmission rate based formulation and performs better than the conventional resource allocation model. Moreover, the compatibility of the proposed model was validated from the QoE and QoS perspectives. Therefore, our proposed semantic-aware resource allocation method could be effective and very useful for the networks where multiple modalities of data, multiple tasks, and multiple systems coexist.
\bibliographystyle{IEEEtran}
\bibliography{IEEEabrv}

\end{document}